\documentclass[aps,pra,twocolumn,groupedaddress]{revtex4-1}

\usepackage{amsmath}
\usepackage{cases}
\usepackage{graphicx}

\usepackage{color}  
\usepackage{enumitem}
\usepackage{multirow}
\usepackage{tabularx}
\usepackage{longtable}
\usepackage{float}

\begin{document}


\title{Green's function coupled cluster formulations utilizing extended inner excitations}

\author{Bo Peng} 
\email{peng398@pnnl.gov}              
\affiliation{William R. Wiley Environmental Molecular Sciences Laboratory, Battelle, 
              Pacific Northwest National Laboratory, K8-91, P.O. Box 999, Richland, WA 99352, USA}
\author{Karol Kowalski} 
\email{karol.kowalski@pnnl.gov}              
\affiliation{William R. Wiley Environmental Molecular Sciences Laboratory, Battelle, 
              Pacific Northwest National Laboratory, K8-91, P.O. Box 999, Richland, WA 99352, USA}


\date{\today}

\begin{abstract}
In this paper we analyze new approximations of the Green's function coupled cluster (GFCC) method where  locations of poles are improved by extending the excitation level of inner auxiliary operators. These new GFCC approximations can be categorized as GFCC-i($n,m$) method, where the excitation level of the inner auxiliary operators ($m$) used to describe the ionization potentials and electron affinities effects in the $N$$-$1 and $N$+1 particle spaces is higher than the excitation level ($n$) used to correlate the ground-state coupled cluster wave function for the $N$-electron system. Furthermore, we reveal the so-called ``$n$+1'' rule in this category (or the GFCC-i($n$,$n$+1) method), which states that in order to maintain size-extensivity of the Green's function matrix elements, the excitation level of inner auxiliary operators $X_p(\omega)$ and $Y_q(\omega)$ cannot exceed $n$+1. We also discuss  the role of the moments of coupled cluster equations that in a natural way assures these properties. Our implementation in the present study is focused on the first approximation in this GFCC category, {\it i.e.} the GFCC-i(2,3) method. As our first practice, we use the GFCC-i(2,3) method to compute the spectral functions for the N$_2$ and CO molecules in the inner and outer valence regimes. In comparison with the GFCCSD results, the computed spectral functions from the GFCC-i(2,3) method exhibit better agreement with the experimental results and other theoretical results, particularly in terms of providing higher resolution of satellite peaks and more accurate relative positions of these satellite peaks with respect to the main peak positions. 
\end{abstract}

\pacs{31.10.+z, 31.15.-p}

\maketitle

\section{Introduction}

Green's function formalism\cite{paldus74_149,paldus75_105,mattuck2012guide,abrikosov2012methods,fetter2012quantum,migdal1967theory}  
is a broadly used tool to calculate properties such as total energies, densities, ionization potentials, electron affinities, and neutral excitations for molecules, clusters, nano-structures, and solids. 
This flexibility of the Green's function formalism, stemming from the possibility of characterizing many-body systems in terms of electron removal and/or addition processes, 
has motivated numerous developments towards providing a theoretical framework that can efficiently combine accuracy and relatively low numerical cost. 
Various computational schemes based on the algebraic and diagrammatic perturbative approaches have been developed so far. 
Typical examples include the outer valence Green's (OVGF) approach,\cite{cederbaum75_290, cederbaum84_57, ortiz97, ortiz13_123} 
the diagonal third-order self-energy approximation (D3) (for the review of diagonal Green function approximations, see Ref. \citenum{zakrzewski2010ab}),
the non-diagonal renormalized second-order approach (NR2),\cite{ortiz98_1008}
the two-hole-one-particle Tamm-Dancoff  approximation ($2h$-$p$ TDA),\cite{cederbaum77_4124}
the third-order quasiparticle method P3,\cite{ortiz96_7599}
the algebraic-diagrammatic construction (ADC),\cite{schirmer82_2395,cederbaum84_57, dreuw15_82} 
and the algebraic approaches based on the intermediate-state representations.\cite{mertins96_2140,mertins96_2153}
In addition, the state-of-the-art $GW$ approximations (one-body Green's function $G$ and dynamically screened Coulomb interaction $W$)\cite{hedin65_a796,rehr00_621,schilfgaarde06_226402,louie11_186404,kas14_085112,reining18_e1344}
permeated many areas of computational chemistry and materials sciences.\cite{onida2002_601,faleev04_126406,louie06_216405}

Recently, one could witness an increasing interest in developing Green's function approaches accounting for higher-order correlation effects, 
which turns out to be indispensable in describing correlated behavior of electrons. 
These efforts serve multiple purposes and have to meet certain requirements. 
For example, in order to support the analysis and interpretation of photoelectron and various X-ray spectroscopies, 
the Green's function formulations have to efficiently deal with various energy regimes. 
Also, as can be seen from a class of Green's function applications that are motivated by the development of various embedding methods including dynamical mean field theories (DMFT)
\cite{kotliar96_13, kotliar06_865, vollhardt12_1, millis06_155107, millis06_076405, zgid11_094115, zgid12_165128} and 
self-energy embedding theory (SEET)\cite{zgid15_121111,zgid15_241102,zgid16_054106,zgid17_2200} approaches, 
very accurate representations of Green's function and/or corresponding self-energies are necessary in describing the so-called impurity regions, which are usually associated with the presence of strong correlation effects.

A hierarchical approximations have been recently proposed to describe self-energies using perturbative techniques for one-particle many-body Green's function (MBGF)\cite{hirata15_1595, hirata17_044108} that have been used to re-derive linked cluster and irreducible-diagram theorems as well as to provide algorithms for general order component of the self-energy. 
A lot of  attention  has also been attracted by the Green's function formulation proposed by Nooijen and Snijders,\cite{nooijen92_55, nooijen93_15, nooijen95_1681}  who successfully employed 
coupled cluster (CC)  bi-orthogonal formalism to express Green's function matrix in terms of the cluster operator ($T$) and the so-called $\Lambda$ operator, which is frequently used in linear response CC theory.\cite{monkhorst77_421,jorgensen90_3333} 
In particular, the latter approach provides a natural elimination of the problems associated with normalization of the ground-state wave function in exponential CC representation. 
More recently, algorithmic developments for calculating Green's function CC (GFCC) has been focused on the design and solving of the auxiliary $\omega$-dependent ionization potential/electron affinity equation-of-motion CC (IP/EA-EOM-CC) type $X_p(\omega)$ and $Y_q(\omega)$ operators acting in the $N$$-$1 and $N$+1 electron spaces (see Refs. \citenum{kowalski14_094102, kowalski16_144101} for details). 
This representation  
can be used to prove/derive basic properties of GFCC including connected character of Green's function matrix elements and their arbitrary-order $\omega$-derivatives.\cite{kowalski16_062512, kowalski18_561} 
The possibility of calculating GFCC matrix and their $\omega$-derivatives, through the Dyson equations, extends in a natural way to the corresponding self-energies. 
These properties enable one to calculate spectral functions, pole strengths, and other properties using GFCC method. 

In this paper we further extend the analysis carried out in the previous papers.\cite{kowalski14_094102, kowalski16_144101,kowalski16_062512, kowalski18_561,kowalski18_4335} 
In particular, motivated by the accuracy attainable by the GFCC model with singles, doubles, and triples (GFCCSDT)\cite{kowalski16_144101} we explore the possibilities of further improving accuracies of calculated pole locations by only increasing excitation level in the auxiliary inner operators $X_p(\omega)$ and $Y_q(\omega)$. 
To this end we discuss the general class of GFCC-i($n,m$) approximations, 
where $n$ is the rank of excitations included in the $T$ and $\Lambda$ operators, 
while $m$ designates the excitation level used to define $X_p(\omega)$ and $Y_q(\omega)$ operators. 
Also, using  consistency conditions between $N$ and $N-1$/$N+1$ particle systems introduced by Nooijen and Snijders\cite{nooijen93_15},
 for the first time ever
we analyze  results of the combined GFCC-i($n$,$m$) formalisms from the point of view of properties of approximate GFCC spectral functions
We will demonstrate that, if $m>n+1$, the disconnected components in the equations for $X_p(\omega)$ and $Y_q(\omega)$ will lead to the appearance of disconnected components in the corresponding Green's function. 
These disconnected components originate in the non-vanishing higher-order moments of the CC equations for $N$-electron ground-state problem.  
Moreover, if $m=n+1$, the equations for $X_p(\omega)$ and $Y_q(\omega)$ operators are defined in terms of connected diagrams, which leads in turn to the fully connected representation of the corresponding GFCC-i($n$,$n$+1) Green's function. 
We will refer to this property of the GFCC-i($n,m$) formulation as an ``$n$+1''-rule. 
We will illustrate the performance of the GFCC-i($n$,$n$+1) formalism by applying the GFCC-i(2,3) method to the computation of the spectral functions of the N$_2$ and CO benchmark systems in the inner and outer valence regime, where several challenging satellite peaks have been identified. 
The GFCC-i(2,3) spectral functions/pole locations  will be compared with 
those obtained with GFCCSD,  IP-EOM-CC,  configuration interaction (CI), and the experimental results. 
We will also discuss the favorable numerical scaling  of the GFCC-i(2,3) method, which obviates the need of correlating the ground state of $N$-electron at the CCSDT level that is associated with very high memory requirements limiting the application area of the GFCCSDT method.


\section{Methodology}

The standard form for the matrix elements of the one-body Green function written in terms of a normalized ground-state wave function $| \Psi \rangle$ for the $N$-electron system can be represented as
\begin{eqnarray}
G_{pq}(\omega) =
&&\langle \Psi | a_q^\dagger (\omega + ( H - E_0 ) - i \eta)^{-1} a_p | \Psi \rangle + \notag \\
&&\langle \Psi | a_p (\omega + ( H - E_0 ) - i \eta)^{-1} a_q^\dagger | \Psi \rangle
\label{gfxn0}
\end{eqnarray}
To construct the GFCC formulation, the direct use of the CC parametrization of the wave function is prohibitive, as in the general case the exponential form of the wave function cannot be easily normalized. To overcome the normalization problem, Nooijen and Snijders\cite{nooijen92_55, nooijen93_15, nooijen95_1681} first employed biorthogonal CC formalism (see for example Refs. \citenum{Schirmer10_145} and \citenum{helgaker2014molecular} for recent reviews of the biorthogonal CC method) to Eq. (\ref{gfxn0}). Similarly, we can also introduce bi-variational CC formalism into Eq. (\ref{gfxn0}). In the bi-variational formalism, the parametrizations of the bra ($\langle \Psi |$) and ket ($| \Psi \rangle$) are defined as follows,
\begin{eqnarray}
\langle \Psi | &=& \langle \Phi | (1+\Lambda)e^{-T}  \label{biv1} \\
| \Psi \rangle &=& e^T | \Phi \rangle.
\label{biv2}
\end{eqnarray}
Plugging Eqs. (\ref{biv1}) and (\ref{biv2}) to Eq. (\ref{gfxn0}), and introducing resolution of identity $1=e^{-T}e^T$, the Green's function CC formulation can then be expressed as
\begin{eqnarray}
G_{pq}(\omega) = 
&&\langle\Phi|(1+\Lambda) \bar{a_q^{\dagger}} (\omega+\bar{H}_N- \text{i} \eta)^{-1} 
	\bar{a}_p |\Phi\rangle + \notag \\
&& \langle\Phi|(1+\Lambda) \bar{a}_p (\omega-\bar{H}_N+ \text{i} \eta)^{-1} 
	\bar{a_q^{\dagger}} |\Phi\rangle \;,
\label{gfxn1}
\end{eqnarray}
where $|\Phi\rangle$ is the reference function, the $\omega$ parameter denotes the frequency, and the imaginary part $\eta$ is often called broadening factor. The similarity transformed operators $\bar{H}_N$ (in its norm product representation), $\bar{a}_p$, and $\bar{a}_q^\dagger$ are defined as
\begin{eqnarray}
\bar{H}_N &=& e^{-T} H ~e^{T} - E_0, \\
\bar{a}_p &=& e^{-T} a_p ~e^{T} = a_p+[a_p,T], \label{shortaim} \\
\bar{a_q^\dagger} &=& e^{-T} a_q^\dagger ~e^{T} = a_q^{\dagger}+[a_q^{\dagger},T] \label{shortaid}.
\end{eqnarray}
Here, $E_0$ is the CC energy, $T$ is the cluster operator, and $\Lambda$ is the de-excitation operator. $T$ and $\Lambda$ are defined as
\begin{eqnarray}
T &=& \sum_{n=1}^{N}
	\frac{1}{(n!)^2}\sum_{\substack{i_1,\ldots,i_n;\\ a_1,\ldots, a_n}} 
	t^{i_1\ldots i_n}_{a_1\ldots a_n} 
	a^{\dagger}_{a_1}\ldots a^{\dagger}_{a_n} 
	a_{i_n}\ldots a_{i_1} \label{tn} \;,\\
\Lambda &=& \sum_{n=1}^{N}
	\frac{1}{(n!)^2}\sum_{\substack{i_1,\ldots,i_n;\\ a_1,\ldots, a_n}} 
	\lambda_{i_1 \ldots i_n}^{a_1\ldots a_n} 
	a^{\dagger}_{i_1} \ldots a^{\dagger}_{i_n} 
	a_{a_n}\ldots a_{a_1} \;,
\end{eqnarray}
with $t^{i_1\ldots i_n}_{a_1\ldots a_n}$ and $\lambda_{i_1 \ldots i_n}^{a_1\ldots a_n}$ being the antisymmetric amplitudes, and the indices $i,j,k,\ldots$ ($i_1,i_2, \ldots$)  and $a,b,c,\ldots$ ($a_1, a_2,\ldots$) corresponding to occupied and unoccupied spin-orbitals in the reference function $|\Phi\rangle$ respectively. $E_0$, $T$, $\Lambda$ can be obtained from the conventional CC equations
\begin{eqnarray}
Q e^{-T}He^T|\Phi\rangle &=& 0 ~, \label{eq:cceq} \\
\langle\Phi|e^{-T}He^T|\Phi\rangle &=& E_0 ~, \label{eeq} \\
\langle\Phi|(1+\Lambda) e^{-T}He^T Q &=& E_0 \langle \Phi|(1+\Lambda)Q ~. \label{eql} 
\end{eqnarray}
where the $Q$ is a projection operator,
\begin{equation}
Q=\sum_{n=1}^N \frac{1}{(n!)^2} \sum_{\substack{i_1,\ldots,i_n;\\ a_1,\ldots, a_n}} |\Phi_{i_1\ldots i_n}^{a_1\ldots a_n}\rangle \langle\Phi_{i_1\ldots i_n}^{a_1\ldots a_n}|\;,
\label{qproj}
\end{equation}
representing the projection onto the subspace spanned by  excited configurations $|\Phi_{i_1\ldots i_n}^{a_1\ldots a_n}\rangle$ defined as  $a_{a_1}^{\dagger} \ldots a_{a_n}^{\dagger}  a_{i_n}\ldots a_{i_1} |\Phi\rangle$. Note that other parametrizations of the $N$-electron wave function can also be used to generate their corresponding Green's function formulations. 

To evaluate the $G_{pq}(\omega)$ matrix elements from Eq. (\ref{gfxn1}), one can formally diagonalize the non-Hermitian effective Hamiltonian $\bar{H}=e^{-T}He^T$ in the ($N\pm1$)-particle space. For the evaluation of the $G_{pq}(\omega)$ matrix elements in the CCSD level, the dimension of the secular matrix ($\bar{H}_N$) is $n_o+n_o^2n_v$ for the retarded part and $n_v+n_v^2n_o$ for the advanced part, where $n_o$ and $n_v$ denote the numbers of occupied molecular orbitals and virtual molecular orbitals, respectively. When systems of interest are getting larger, the increasing number of orbitals prohibits the direct diagonalization of the secular matrix, and iterative algorithm such as {\it Krylov} subspace methods are often used. However, since only part of the spectrum is solved (or converged) by the 
conventional
iterative subspace diagonalization methods, the introduction of the incomplete eigenpairs to the GFCC formulation would lead to a sum-over-states representation of the GFCC formulation (see Ref. \citenum{kowalski16_144101}), which is usually unfavored in a sense that there is unclearness of how many states should be involved in the representation. Besides, for systems that exhibit a significant many-body effect in the spectral region of interest, solving the secular equations by 
the conventional
iterative diagonalization is often subject to nontrivial construction of initial vectors and convergence problems. 
To circumvent these problems, approximations such as core-valence separation (CVS)\cite{cederbaum80_206} has been introduced in the high-order Green's function method to compute K-shell ionization spectra of small and medium-size molecules\cite{schirmer87_6789, dreuw14_4583}. 
Essentially, the CVS approximation neglects the coupling between core- and valence-excited states (zeroes certain types of Coulomb integrals in practice), and therefore, reduces the dimension of the effective Hamiltonian for the problem of interest. The reduction of dimension of the effective Hamiltonian becomes more significant for core-electron ionization than for the valence-electron ionization, and the associated error was estimated to be 0.5$\sim$1.0 eV for the former, and much smaller for the latter\cite{trofimov00_483}. In particular, for the core-electron ionization, the performance of the convergence would be greatly improved when the CVS is employed, and vast majority states of interest would then be easily positioned. In the context of wave-function-based or propagator-based $ab$ initio methods, the CVS approximation has been applied to the ADC, CC2, CCSD, CC3, and CCSDR(3) methods (see Ref. \citenum{norman18_7208} for a recent review). Recently, on the computational side, more algorithms have also been proposed for iterative matrix-free eigensolvers
(such as asymmetric Lanczos-chain-driven subspace algorithm\cite{coriani12_1616}, energy-specific Davidson algorithm\cite{peng15_4146}, and Generalized Preconditioned Locally Harmonic Residual (GPLHR) method\cite{zuev14_273}, etc.) have been proposed to work with linear-response coupled-cluster (LR-CC) or EOM-CC method to solve not only the lowest/highest eigenvalues but also the inner eigenvalues embedded deeply in the spectrum of the large Jacobian matrix, such that accurate resolution can be obtained for the frequency window of interest, and viable approximation can even be generated for the entire spectrum. For the GFCC calculation, as mentioned in a recent study of the optical potential calculation for the nuleon-nucleon scattering\cite{rotureau17_024315}, the advantage of using Lanczos-based method lies in the fact that the resolution of the tridiagonal matrix representing the normal-ordered Hamiltonian matrix in the Lanczos basis only needs to be done once for all frequencies ($\omega$'s).
Alternatively, we recently proposed to first solve a set of linear equations for IP/EA-EOMCC type vectors (using their zero-th and first order terms as starting points), and then to contract these vectors with the converged amplitudes of the CC $\Lambda$ de-excitation operators to get the GFCC matrix elements.\cite{kowalski14_094102, kowalski16_144101} The computational approach was designed to calculate GFCC for the whole complex plane, therefore it includes all poles of the GFCC structure, and serves naturally for, for example, an embedding scheme when working with low-level methods for large-scale applications. Facilitated by this approach, we then were able to prove the connected character of the diagrams contributing to GFCC matrix elements, as well as the connected character of its $n$-th order derivative with respect to the energy and the corresponding CC self-energy operators,\cite{kowalski16_062512, kowalski18_561} which provided a useful guidance for designing and analyzing new and size-extensive GFCC approximation schemes. Furthermore, due to its algebraic structure, the proposed method is highly scalable, and is capable of computing spectral function for a given molecular system in any energy region. We recently demonstrated this capability for several typical molecular systems (such as H$_2$O, N$_2$, CO, 1,3-butadiene, benzene, adenine molecules)\cite{kowalski18_4335}. Consistent with previous ADC results\cite{schirmer05_144115,deleuze06_104309,ortiz09_14630,schirmer08_360}, satellite peaks have been observed from the computed spectral functions within the GFCCSD framework in the energy regions where many-body effect becomes significant and single particle picture of ionization often breaks down\cite{cederbaum77_L549,cederbaum80_481}. 
It should be emphasized here that solving a set of linear equations and diagonalizing secular matrix are purely two different computational procedures with similar scaling (i.e. $\mathcal{O}(N^6)$ for CCSD), and different computational procedure in the same GFCC theoretical context will not add anything new to the theory itself. Therefore, the peak positions obtained from the two computational procedures should be exactly same within the same truncated subspace. Also, It needs to point out that  the CVS approximation mentioned above is in principle applicable to either a diagonalization routine or a linear solver since similar tensor contractions need to be performed in both routines (see $\bar{H}_N X_p$ in the equations below). Nevertheless, it should be pointed out (also see the discussion in the results section), the disadvantage of solving linear equations in comparison to the diagonalization routine lies in that facts that (i) a much smaller frequency interval is necessary in order to explore some detailed information (in particular the state of interest with weak intensity) which unavoidably increases the computation cost, and (ii) it fails to identify any dark state. Despite of these disadvantages, solving linear equations has been chosen by some other studies recently for the computation of the spectral function of,
for example, uniform electron gas\cite{chan16_235139}, light atoms\cite{matsushita18_034106}, heavy metal atoms\cite{matsushita18_224103}, and simple 1-D periodic systems\cite{matsushita18_204109}.
A key step in our procedure is to, in a computationally tractable way, introduce an $\omega$-dependent IP/EA-EOMCC type operators $X_p(\omega)$ in an ($N$$-$1)-electron Hilbert space and an $\omega$-dependent EA-EOMCC type operators $Y_q(\omega)$ in an ($N$+1)-electron Hilbert space
\begin{eqnarray}
X_p(\omega) &=& \sum_{i} x^i(p, \omega)  a_i  + \sum_{i<j,a} x^{ij}_a(p, \omega) a_a^{\dagger} a_j a_i +\ldots ~, \label{xp} \notag \\ \\
Y_q(\omega) &=& \sum_{i} y^a(q, \omega) a_a^\dagger  + \sum_{i,a<b} y^{i}_{ab}(q, \omega) a_a^{\dagger} a_b^\dagger a_i +\ldots ~, \label{yp} \notag \\ 
\end{eqnarray}
which satisfy 
\begin{eqnarray}
(\omega+\bar{H}_N - \text{i} \eta )X_p(\omega)|\Phi\rangle = 
	\bar{a}_p |\Phi\rangle \;, \label{eq:xplin} \\
(\omega-\bar{H}_N + \text{i} \eta )Y_q(\omega)|\Phi\rangle = 
	\bar{a_q^\dagger} |\Phi\rangle \;,\label{eq:yqlin}
\end{eqnarray}
and can be used to write a compact expression for the GFCC 
\begin{eqnarray}
G_{pq}(\omega) = 
\langle\Phi|(1+\Lambda) \bar{a_q^{\dagger}} X_p(\omega) |\Phi\rangle + \langle\Phi|(1+\Lambda) \bar{a}_p Y_q(\omega) |\Phi\rangle \;. \notag \\
\label{gfxn2}
\end{eqnarray}
%
%

%
\begin{figure}
\includegraphics[angle=270, width=0.5\textwidth]{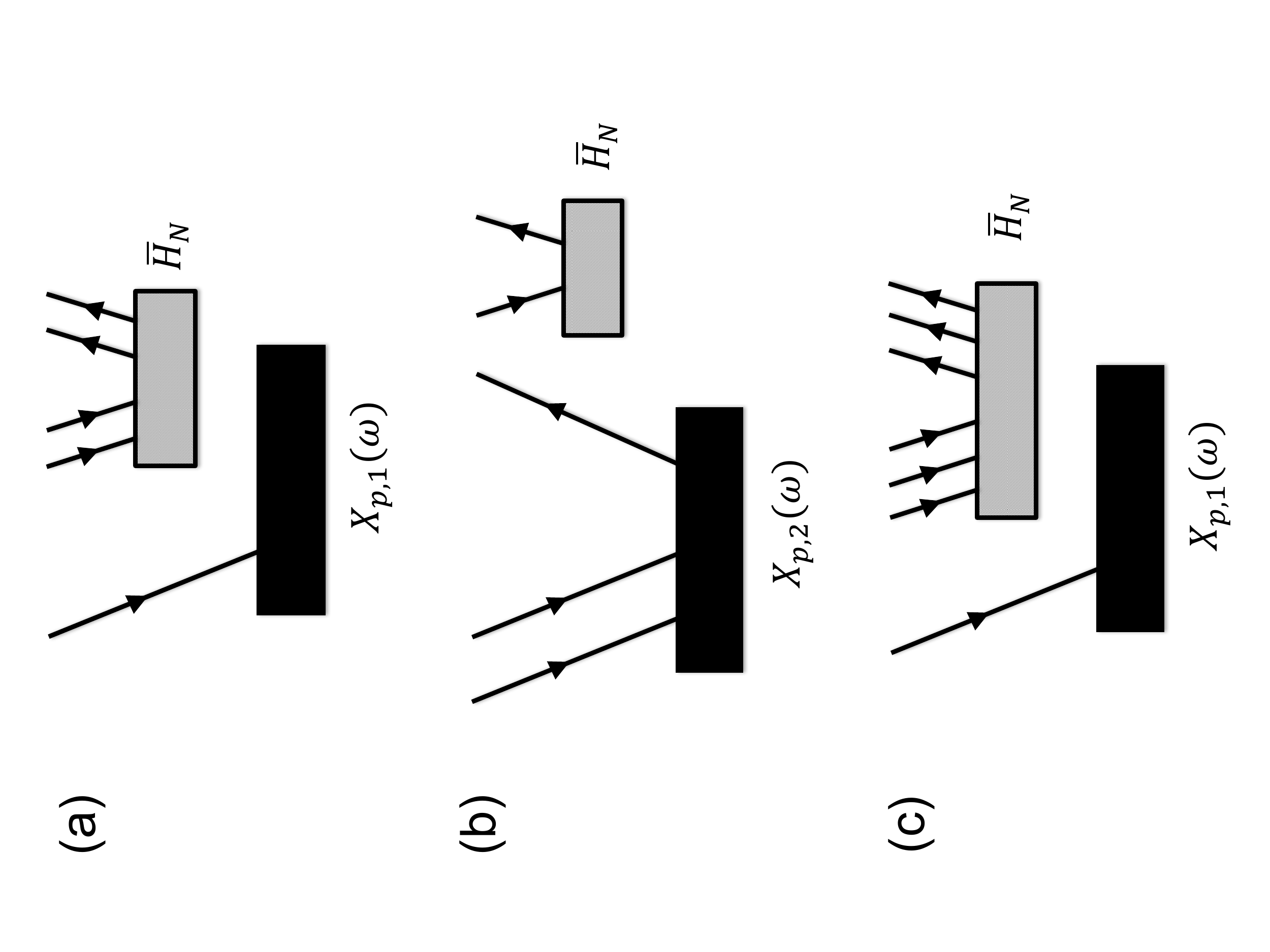}
\caption{Disconnected contributions in the equations for triples ($X_{p,3}(\omega)$) in the GFCC-i(2,3) method (insets (a) and (b)) and in the equations for quadruples ($X_{p,4}(\omega)$) in the GFCC-i(2,4) method (inset (c)).}
\label{itcon}
\end{figure}

Typical approximations in the GFCC formulation discussed in early papers\cite{kowalski14_094102, kowalski16_144101,kowalski16_062512, kowalski18_561,kowalski18_4335} are defined by limiting the rank of excitation included in the all $T$, $\Lambda$, $X_p(\omega)$ and $Y_q(\omega)$ operators. 
This general approximation procedure determines the class of correlation effects used to describe the ground state of $N$-electron system but also 
impacts the quality of pole locations described by $X_p(\omega)$ and $Y_q(\omega)$ for ($N$$-$1)- and ($N$+1)-electron systems. 
For example, in the GFCCSD approximation (GFCC with singles and doubles), 
the expansions for these operators are  truncated at singles and doubles level, {\it i.e.},
\begin{eqnarray}
T &\approx& \sum_{i,a} t_a^i a_a^\dagger a_i + 
	\sum_{i<j,a<b}  t_{ab}^{ij} a_a^\dagger a_b^\dagger a_j a_i ~, \label{t1t2} \\
\Lambda &\approx& \sum_{i,a} \lambda_i^a a_i^\dagger a_a +
	\sum_{i<j,a<b} \lambda_{ij}^{ab} a_i^\dagger a_j^\dagger a_b a_a ~, \label{l1l2} \\
X_p(\omega) &\approx& \sum_{i} x^i(p, \omega)  a_i  + 
	\sum_{i<j,a} x^{ij}_a(p, \omega) a_a^{\dagger} a_j a_i ~, \label{xpsd}\\
Y_q(\omega) &\approx& \sum_{i} y^a(q, \omega) a_a^\dagger  + 
	\sum_{i,a<b} y^{i}_{ab}(q, \omega) a_a^{\dagger} a_b^\dagger a_i ~. \label{yqsd}
\end{eqnarray}
and  $X_p(\omega)$ and $Y_q(\omega)$ operators are determined from 
\begin{widetext}
\begin{align} 
&&(Q_1^{(N-1)}+Q_2^{(N-1)})(\omega+\bar{H}_N - \text{i} \eta ) (Q_1^{(N-1)}+Q_2^{(N-1)})
X_p(\omega)|\Phi\rangle 
= 
(Q_1^{(N-1)}+Q_2^{(N-1)})	\bar{a}_p |\Phi\rangle  \;, \label{eq:xp} \\
&&(Q_1^{(N+1)}+Q_2^{(N+1)})  (\omega-\bar{H}_N + \text{i} \eta ) (Q_1^{(N+1)}+Q_2^{(N+1)})
Y_q(\omega)|\Phi\rangle  
= 
(Q_1^{(N+1)}+Q_2^{(N+1)})	\bar{a_q^\dagger} |\Phi\rangle \;.\label{eq:yq}
\end{align} 
\end{widetext}
where  subspaces on which Eqs.(\ref{eq:xp}) and (\ref{eq:yq}) are projected on 
are 
chosen to reflect excitation included in the CCSD $X_p(\omega)$ and $Y_q(\omega)$ operators. This leads to the following 
for the CCSD Green's function: 
\begin{eqnarray}
G_{pq}(\omega) &=&  
\langle\Phi|(1+\Lambda_1+\Lambda_2) \bar{a_q^{\dagger}} (X_{p,1}(\omega)+X_{p,2}(\omega)) |\Phi\rangle \;, \nonumber \\
&& + \langle\Phi|(1+\Lambda_1+\Lambda_2) \bar{a_p} (Y_{q,1}(\omega)+Y_{q,2}(\omega)) |\Phi\rangle \notag \\
\label{gfxn2}
\end{eqnarray}
which has been used in several studies to determine efficiency of the GFCCSD approximations. 

As discussed in Ref.\cite{kowalski16_144101}, GFCC formalism offers a flexibility in treating correlation effect. 
In particular, different  levels of approximations can be used to describe ground-state effects for the $N$-electron systems ($T$ and $\Lambda$ operators)  and location of poles for $N$$-$1 and $N$+1 systems ($X_p(\omega)$ and $Y_q(\omega)$ operators). 
In this paper, we will establish a simple rule that assures that the basic properties of the GFCC formalism, such as connected character of contributing diagrams,  are still preserved.  
We will refer to these schemes as GFCC-i($n,m$) where $n$ and $m$ define the level of excitations used to define approximate operators $T$/$\Lambda$ and $X_p(\omega)$/$Y_q(\omega)$, respectively, 
and `i' stands for the inner space indicating that two-levels of approximations are used. 
An interesting feature of the GFCC-i($n,m$) is the fact that, when 
$m=N$, the $X_p(\omega)$ and $Y_q(\omega)$ operators reproduce exact location of poles in the $N$$-$1 and $N$+1 spaces.
This is a straightforward consequence that similarity transformation $e^{-T}He^T$ does not change the spectral properties of the electronic Hamiltonian $H$. 

As a specific example, let us focus on the GFCC-i(2,3) method. 
In this case, in addition to singly ($X_{p,1}(\omega$)/$Y_{q,1}(\omega)$) and doubly ($X_{p,2}(\omega$)/$Y_{q,2}(\omega)$) excited components of $X_p(\omega)$ and $Y_q(\omega)$ operators we add three-body terms ($X_{p,3}(\omega$)/$Y_{q,3}(\omega)$), which leads to the following definitions of these operators 
\begin{widetext}
\begin{align} 
&&X_p(\omega) \approx \sum_{i} x^i(p, \omega)  a_i  + 
	\sum_{i<j,a} x^{ij}_a(p, \omega) a_a^{\dagger} a_j a_i + 
	\sum_{i<j<k,a<b} x^{ijk}_{ab} (p, \omega) a_a^{\dagger} a_b^{\dagger} a_k a_j a_i ~, \label{xpsdt}\\
&&Y_q(\omega) \approx \sum_{i} y^a(q, \omega) a_a^\dagger  + 
	\sum_{i,a<b} y^{i}_{ab}(q, \omega) a_a^{\dagger} a_b^\dagger a_i +
        \sum_{i<j,a<b<c} y^{ij}_{abc}(q, \omega) a_a^{\dagger} a_b^\dagger a_c^\dagger a_j a_i 	
	~. \label{yqsdt}
\end{align}
\end{widetext}
At the same time,  the $T$ and $\Lambda$ operators in the GFCC-i(2,3) are given by Eqs. (\ref{t1t2}) and (\ref{l1l2}). 
The working equations for the $X_p(\omega)$ and $Y_q(\omega)$ in the GFCC-i(2,3) method are given by expressions
\begin{widetext}
\begin{align}
(Q_1^{(N-1)}+Q_2^{(N-1)}+Q_3^{(N-1)})(\omega+\bar{H}_N - \text{i} \eta ) 
(X_{p,1}(\omega)+X_{p,2}(\omega)+X_{p,3}(\omega))|\Phi\rangle 
= (Q_1^{(N-1)}+Q_2^{(N-1)}+Q_3^{(N-1)})	\bar{a}_p |\Phi\rangle \;, \label{eq:xp23} \\
(Q_1^{(N+1)}+Q_2^{(N+1)}+Q_3^{(N+1)})  (\omega-\bar{H}_N + \text{i} \eta ) 
(Y_{q,1}(\omega)+Y_{q,2}(\omega)+Y_{q,3}(\omega))|\Phi\rangle 
=(Q_1^{(N+1)}+Q_2^{(N+1)}+Q_3^{(N+1)})  \bar{a_q^\dagger} |\Phi\rangle  \;.\label{eq:yq23}
\end{align}
\end{widetext}
The key role in assuring the connected character of the CC Green's function matrix elements was played by the fact that the $X_p(\omega)$ and $Y_q(\omega)$ operators are expressible in terms of connected diagrams. 
In Refs.\cite{kowalski16_062512,kowalski18_561}, we demonstrated that this feature is a consequence of the connected character of equations for the $X_p(\omega)$ and $Y_q(\omega)$ operators. 
It is interesting to analyze this property from the point of view of Eqs.(\ref{eq:xp23}) and (\ref{eq:yq23}).  
By closer inspection (see Fig. \ref{itcon}a,b) one can notice that the only disconnected diagrams contributing to Eq. (\ref{eq:xp23}) correspond to diagrams that involve elements of $\bar{H}_N$ corresponding to CCSD equations (examples on these diagrams corresponding to projections on triples are also shown in Fig. \ref{itcon}a,b, analogous behavior can also be found for the $Y_q(\omega)$ equations), and numerically disappear once the CCSD ground state converges. 
Therefore, the equations for $X_p(\omega)$ and $Y_q(\omega)$ GFCC-i(2,3) amplitudes ({\it i.e.} Eqs. (\ref{eq:xp23}) and (\ref{eq:yq23})) involve only connected diagrams. 
Given the fact that  GFCC-i(2,3)  matrix elements are given by the same expression, Eq. (\ref{gfxn2}), as in the GFCCSD case (with the difference that in the GFCC-i(2,3) case $X_{p,i}(\omega)$ and $Y_{q,i}(\omega)$ ($i=1,2$) are iterated in the presence of  triply excited amplitudes $X_{p.3}(\omega)$ and $Y_{q,3}(\omega)$), their diagrammatic expansion contains connected diagrams only. 

\begin{figure}
\includegraphics[angle=270,width=0.5\textwidth]{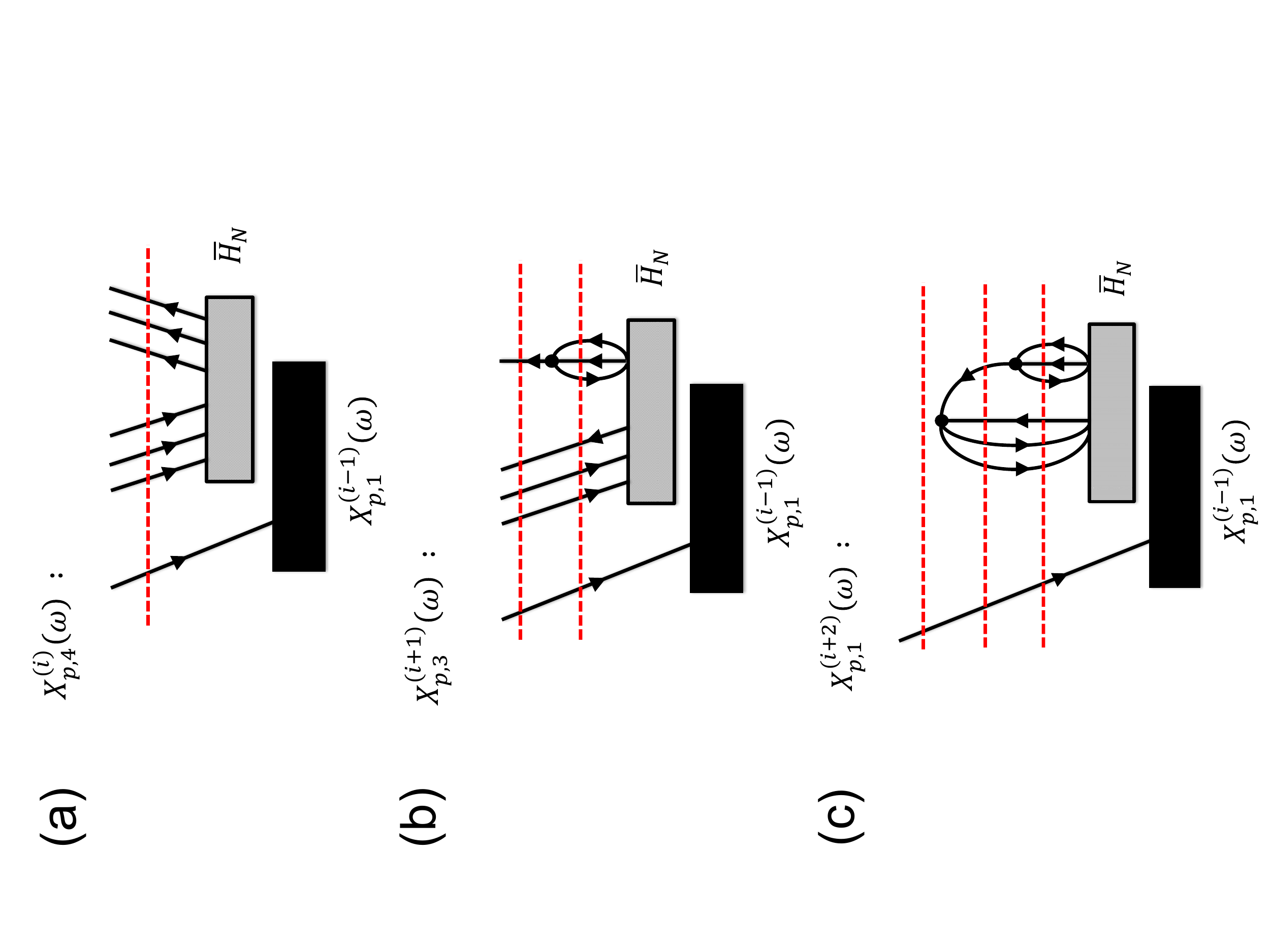}
\caption{The mechanism of formation of disconnected contributions in the GFCC-i(2,4) formalism: (a) Disconnected yet linked contribution to the $X_{p,4}(\omega)$ operator in the $i$-th iterative cycle. 
(b) Disconnected yet linked contribution to the $X_{p,3}(\omega)$ operator in the ($i$+1)-th iteration. 
Diagram shown in step (b) refers to typical $Q_3^{(N-1)}V_N X_{p,4}(\omega)|\Phi\rangle$  ($V_N$ refers to the two-body part of electronic Hamiltonian in normal product form).
(c) Disconnected diagram (containing disconnected closed part) contributing to $X_{p,1}(\omega)$  in the ($i$+2)-th iteration through the coupling term $Q_1^{(N-1)} V_N X_{p,3}(\omega)|\Phi\rangle$.
Red dotted lines represent symbolically zeroth order resolvents.}
\label{itdis}
\end{figure}

However, one should proceed with caution when extending GFCC-i(2,3) to higher $m$'s. 
For example, in the GFCC-i(2,4) a disconnected diagram shown in Fig. \ref{itcon}c appears in the equations for quadruples ($X_{p,4}(\omega)$ equations), which, as explained in Fig. \ref{itdis}, gives rise  during the iterations to a disconnected components entering $X_p(\omega)$ operators that correspond to lower order excitations. 
The provenance of these problems is associated with the non-vanishing character of higher-order moments of the CCSD equations (see Refs. \cite{Jankowski91_223, kowalski00_18}), which are not used to define CCSD equations.  
Therefore, the only GFCC-i($n$,$m$) methods that guarantee the connectivity of the corresponding matrix elements of the Green's function are of the GFCC-i($n$,$n$+1) type.

Generally speaking, the  GFCC-i($n$,$m$) formalism can be viewed as an extension of previous studies of approximate inclusion of 
higher-order excitations in the IP/EA-EOMCC methods represented by a broad class of iterative and non-iterative methodologies including 
EOMIP-CCSD$^{\ast}$\cite{stanton1999simple,saeh1999application,manohar2009perturbative},
CCSDT-n (n=1,2,3)
\cite{bartlett85_4041,bartlett95_81,bartlett96_581,stanton99_8785,bomble2005equation}, 
CCn (n=2,3)\cite{koch97_1808, christiansen95_409},  EOM-CC(m,n)\cite{hirata00_255,hirata06_074111,krylov05_084107}, and 
EOM-CCSD(T)(a)$^{\ast}$  formalisms\cite{matthews2016new,jagau2018non}.
Moreover, the EOM-CC(m,n) approach introduced by Hirata {et al.} combines  different levels of approximations for the ground-state CC Ansatz  and equation-of-motion CC operators $R_K$  for vertical excitation  energies, ionization potential, and electron affinities has already been applied to molecular systems. 
Several problems of the EOM-CC(m,n) class of methods stemming from possible inconsistencies between  approximation levels of the $T$ and $R_K$ operators approaches 
and resulting in the lack of the  size-intensivity of the calculated excitation energies, have been discussed by Splipchenko and Krylov\cite{krylov05_084107}.
In this paper, for the first time 
we demonstrate that similar procedures applied to $X_p(\omega)$ and $Y_q(\omega)$ operators (which in contrast to the $R_K$ operators 
are determined by solving linear equations) may   violate size-extensive character of the corresponding Green's function if 
difference between excitation levels used to approximate $X_p(\omega)$/$Y_q(\omega)$ operators and $T$ operators are bigger than one. 

In this paper, we will focus on the GFCC-i(2,3) method which introduces triple excitations in describing correlation effects needed for the accurate determination of the poles of the Green's function. 
In contrast to the GFCCSDT method, the GFCC-i(2,3) formalism bypasses memory demands for storing $T_3$ amplitudes (proportional to $n_o^3 n_v^3$, where $n_o$ and $n_v$ refer to the total number of occupied and virtual orbitals, respectively), and the maximum memory demand in the GFCC-i(2,3) formalism comes from storing $X_{p,3}(\omega)$ ($Y_{q,3}(\omega)$) amplitudes that is proportional to $n_o^3 n_v^2$ ($n_o^2 n_v^3$).

To obtain the working equations for the GFCC-i(2,3) method, we have chosen a form of Eq. (\ref{eq:xp23}), in which the projections on singles and doubles are same as in GFCCSDT (see Eqs. (\ref{x3_q1}) and (\ref{x3_q2}) where the full form of $X_{p,3}(\omega)$-dependent coupling terms including $Q_1^{(N-1)}V_N X_{p,3}(\omega)|\Phi\rangle$, $Q_2^{(N-1)}V_N X_{p,3}(\omega)|\Phi\rangle$, and $Q_2^{(N-1)}(V_N T_1 X_{p,3}(\omega))_C|\Phi\rangle$ is maintained).
\begin{widetext}
\begin{align}
&&Q_1^{(N-1)}(\omega+\bar{H}_N - \text{i} \eta ) 
(X_{p,1}(\omega)+X_{p,2}(\omega)+X_{p,3}(\omega))|\Phi\rangle = Q_1^{(N-1)} \bar{a_p}|\Phi\rangle~, \label{x3_q1} \\
&&Q_2^{(N-1)}(\omega+\bar{H}_N - \text{i} \eta ) 
(X_{p,1}(\omega)+X_{p,2}(\omega)+X_{p,3}(\omega))|\Phi\rangle = Q_2^{(N-1)} \bar{a_p}|\Phi\rangle~. \label{x3_q2}
\end{align}
\end{widetext}
As to the projections on triples, in contrast to GFCCSDT, we chose to maintain the leading contributing $X_{p,1}$, $X_{p,2}$, and $X_{p,3}$ operators, {\it i.e.}

\begin{widetext}
\begin{align}
Q_3^{(N-1)} \lbrace ((\omega+F_N- \text{i} \eta) X_{p,3}(\omega))_C  + (V_N X_{p,2}(\omega))_C + 
(V_N T^{(1)}_2 X_{p,1}(\omega))_C \rbrace  |\Phi\rangle 
\approx Q_3^{(N-1)} \bar{a_p}|\Phi\rangle  \;,  \notag \\ \label{x3_q3}
\end{align}
\end{widetext}
where the one-particle part $F_N = \sum_{r} \epsilon_r N[a_r^\dagger a_r ]$, and the two-particle part $V_N = \frac{1}{4}\sum_{p,q,r,s}$ $v_{pq}^{rs} N[a_p^\dagger a_q^\dagger a_s a_r]$ with $v_{pq}^{rs}$ denoting antisymmetrized two-electron integrals and $N[\ldots]$ designating normal ordered form of a given second-quantized expression.

Remarkably, analogous to CCSDT-n (n=1,2,3) and  CC3 methods, in Eq. (\ref{x3_q3}), the triply excited component $X_{p,3}(\omega)$ only contracts with zero-th order terms, 
i.e., the $3h2p$-$3h2p$ coupling is only represented by the lowest-order contribution stemming from the Fock operator (note that full inclusion of $\bar{H}_N$ in the $3h2p$-$3h2p$ coupling would significantly increase the numerical overhead from $\mathcal{O}(N^6)$ to $\mathcal{O}(N^7)$ per iteration).
 which allows the on-the-fly determination of $X_{p,3}(\omega)$  as a function of $X_{p,1}(\omega)$ and $X_{p,2}(\omega)$ (hereafter, we call $X_{p,3}(\omega)$ of this type internal $X_{p,3}(\omega)$, or internal triple). That is, based on Eq. (\ref{x3_q3}), the components of $X_{p,3}(\omega)$, $x^{ijk}_{ab,p}(\omega)$ (including both the the real and imaginary parts), can be obtained through
\begin{widetext}
\begin{align}
&&\Re\left(x^{ijk}_{ab,p}(\omega)\right) =
\frac {1} {\Delta \epsilon^{ijk}_{ab}(\omega)^2 + \eta^2}
\left[ \Delta \epsilon^{ijk}_{ab}(\omega) ~\Re\left(U^{ijk}_{ab, p}(\omega)\right) - \eta~ \Im\left(U^{ijk}_{ab, p}(\omega)\right) \right] ~, \label{x3_r} \\
&&\Im\left(x^{ijk}_{ab,p}(\omega)\right) =
\frac {1} {\Delta \epsilon^{ijk}_{ab}(\omega)^2 + \eta^2}
\left[ \Delta \epsilon^{ijk}_{ab}(\omega) ~\Im\left(U^{ijk}_{ab, p}(\omega)\right) + \eta~ \Re\left(U^{ijk}_{ab, p}(\omega)\right) \right] ~, \label{x3_i}
\end{align}
\end{widetext}
where 
\begin{widetext}
\begin{align}
&&U^{ijk}_{ab, p}(\omega) = 
A_1 \{ v^{ij}_{la} x^{lk}_{b, p}(\omega) - t^{ca}_{jk} v^{db}_{ic} x_{d, p}(\omega) \} - 
A_2 \{ v^{ab}_{ic} x^{jk}_{c, p}(\omega) + t^{ab}_{lk} v^{cl}_{ij} x_{c, p}(\omega) \} ~,  \label{U} 
\end{align}
\end{widetext}
and $\Delta \epsilon^{ijk}_{ab}(\omega) = \omega + \epsilon_a + \epsilon_b - \epsilon_i - \epsilon_j - \epsilon_k$ with $A_1\{...\}$ and $A_2\{...\}$ representing corresponding antisymmetric permutation operators, and $\epsilon_r$ denoting the $r$-th Hartree-Fock orbital energy. In the above expressions, $i,j,k,l,\cdots$ denote occupied Hartree-Fock orbital indices, $a,b,c,d,\cdots$ denote virtual Hartree-Fock orbital indices, and $p,q,r,s,\cdots$ denote general Hartree-Fock orbital indices. 
Note that the computational cost of Eqs. (\ref{x3_r}) and (\ref{x3_i}), for a given HF orbital $p$ and frequency $\omega$, only scales as $n_o^3 n_v^2$, while the cost of Eq. (\ref{U}) scales as $n_o^3 n_v^3$. 

%


Once $X_{p,3}(\omega)$ is generated, the major goal is then to solve Eqs. (\ref{x3_q1}) and (\ref{x3_q2}), which can alternatively be represented using the following compact form
\begin{equation}
(\omega + \bar{\bf{H}}_N - \text{i} \eta) {\bf x}_p(\omega)
	= {\bf b}_p ~, \label{mat1} 
\end{equation}
in which the index $p$ runs over the entire spin-orbital domain, and ${\bf b}_p$ is a real free term. As discussed in our previous work, by representing ${\bf x}_p(\omega)$ as a vector
\begin{equation}
{\bf x}_p(\omega) = 
\left[ \begin{array}{c} 
\Re \left( {\bf x}_p(\omega) \right)\\ 
\Im \left( {\bf x}_p(\omega) \right)
\end{array} \right]
~,
\end{equation}
then the matrix form (\ref{mat1}) becomes
\begin{eqnarray}
\left[ \begin{array}{cc}
\omega + \bar{\bf{H}}_N & \eta \\ -\eta & \omega + \bar{\bf{H}}_N \end{array} \right]
\left[ \begin{array}{c} 
\Re \left( {\bf x}_p(\omega) \right)\\ 
\Im \left( {\bf x}_p(\omega) \right)
\end{array} \right]
=
\left[ \begin{array}{c} 
{\bf b}_p\\ 
{\bf 0}
\end{array} \right]. \label{newlineq}
\end{eqnarray}
Eq. (\ref{newlineq}) can be approximately solved by iterative procedures ({\it i.e.} linear equation solvers). Once the one-body and two-body parts of ${\bf x}_p(\omega)$ is converged, the GFCC matrix element can then be obtained through Eq. (\ref{gfxn2}).

The overall procedure of executing a GFCC-i(2,3) calculation for the retarded part of $G(\omega)$ can then be summarized as three parts, {\it i.e.}, ($\Lambda$-)CCSD calculations that scale as $n_o^2 n_v^4$, GFCC-i(2,3) iterative part in which each iteration step scales as $n_o^3 n_v^3$ iteration for a given pair of $p$ and $\omega$, and one-shot contraction (Eq. (\ref{gfxn2})) that scales as $n_o^2 n_v^2$ for a given pair of $p$ and $\omega$. It is worth mentioning that, in comparison with the GFCCSDT method, the GFCC-i(2,3) method doesn't require the cumbersome $n_o^3 n_v^5$ ($\Lambda$-)CCSDT calculations. Furthermore, in spite of having the same numerical scalings during the iteration, the iteration part in the GFCC-i(2,3) method bears a much smaller prefactor than that in the GFCCSDT method. Also, in terms of storage requirement, the GFCC-i(2,3) method eases off the storage of $X_{p,3}(\omega)$ at the end of the iteration. Besides, in general, due to the inherent independencies in the GFCC formalism ({\it i.e.} the calculation of $X_p(\omega)$ for a given pair of $p$ and $\omega$ is independent of the calculation for another pair), our GFCC method is a strong scaling method that  allows the efficient utilization of process groups in the calculation to significantly reduce time-to-solution, and is suitable for the simulation of larger systems with large basis sets.\cite{kowalski14_094102}

As above mentioned, most of the computation cost in the GFCC-i(2,3) calculation is spent on solving the linear equation, Eq. (\ref{newlineq}), which is closely related to the linear equation solvers used in the calculation. In the present study, we chose to solve Eq. (\ref{newlineq}) by exploiting the so-called direct inversion in the iterative subspace (DIIS) procedure. Different from conventional DIIS approach, here we adapt the DIIS to the complex space. Supposed we have $K$ micro-iteration iterates ${\bf x}_{p,j+1}(\omega)$, $\cdots$, ${\bf x}_{p,j+K}(\omega)$, then the new ${\bf x}_{p,j+K+1}(\omega)$ is obtained through a linear combination of the $K$ micro-iteration iterates
%
\begin{eqnarray}
{\bf x}_{p,j+K+1}(\omega) 
&=& \sum_{k=1}^K c_k \tilde{\bf x}_{p,j+k+1}(\omega) \notag \\
&=& \sum_{k=1}^K c_k ( {\bf x}_{p,j+k}(\omega) + {\bf r}_{p,j+k}(\omega) ) \label{xupdate}
\end{eqnarray}
%
where $\sum_{k=1}^K c_k = 1$, and ${\bf r}_{p,j+k}(\omega)$ represents the residual vector of ${\bf x}_{p,j+k}(\omega)$, and $\tilde{\bf x}_{p,j+k+1}(\omega)$ is an intermediate update of ${\bf x}_{p,j+k}(\omega)$, and the coefficient vector ${\bf c} = \left[ c_1, \cdots, c_k, \cdots, c_K \right]^T$ is defined in the complex space. The goal is to minimize the DIIS least square functional $J_{\text{DIIS}} := \displaystyle \| \sum_{k=1}^K c_k{\bf r}_{p,j+k} \|_2$ by solving the following system of linear equations
\begin{equation}
\left[ \begin{array}{cc} 
{\bf B} &  {\bf -1} \\ 
{\bf -1}^T &  0
\end{array} \right] \cdot
\left[ \begin{array}{c} 
{\bf c} \\ 
\zeta 
\end{array} \right] =
\left[ \begin{array}{c} 
{\bf 0} \\ 
-1
\end{array} \right]
~, \label{diis}
\end{equation}
where ${\bf 1} = \left[ 1, \cdots, 1 \right]$ is a constant vector of length $K$, $\zeta$ is a complex {\it Lagrangian} multiplier, and the $m\times m$ matrix ${\bf B}$ is constructed through,
\begin{equation}
{\bf B} = 
\left[ \begin{array}{c} 
{\bf r}_{p,j+1}^T(\omega) \\ \vdots \\
{\bf r}_{p,j+K}^T(\omega) 
\end{array} \right] \cdot
\left[ \begin{array}{ccc}
{\bf r}_{p,j+1}(\omega) & \ldots & {\bf r}_{p,j+K}(\omega)
\end{array} \right]
\end{equation}
If we assume the $\bar{\bf{H}}_N$ matrix is diagonal dominant, the correction vector at the ($j$+$k$)-th step, ${\bf r}_{p,j+k}(\omega)$, can be approximated through
%
\begin{eqnarray}
\left[ \begin{array}{c} 
\Re \left( {\bf r}_{p,j+k}(\omega) \right)\\ 
\Im \left( {\bf r}_{p,j+k}(\omega) \right)
\end{array} \right]
\approx 
\left[ \begin{array}{cc}
\omega + \bf{D} & {\bf \eta} \\ {\bf -\eta} & \omega + \bf{D} \end{array} \right]^{-1}
\left\{
\left[ \begin{array}{c} 
{\bf b}_p\\ 
{\bf 0}
\end{array} \right] \right. \notag\\
\left. -
\left[ \begin{array}{cc}
\omega + \bar{\bf{H}}_N & \eta \\ -\eta & \omega + \bar{\bf{H}}_N \end{array} \right]
\left[ \begin{array}{c} 
\Re \left( {\bf x}_{p,j+k}(\omega) \right)\\ 
\Im \left( {\bf x}_{p,j+k}(\omega) \right)
\end{array} \right]
\right\}
\end{eqnarray}
%
with $\bf{D}$ being the orbital-energy dependent diagonal part  of  the $\bar{\bf{H}}_N$ matrix. 

\begin{figure*}
\includegraphics[width=1.0\textwidth]{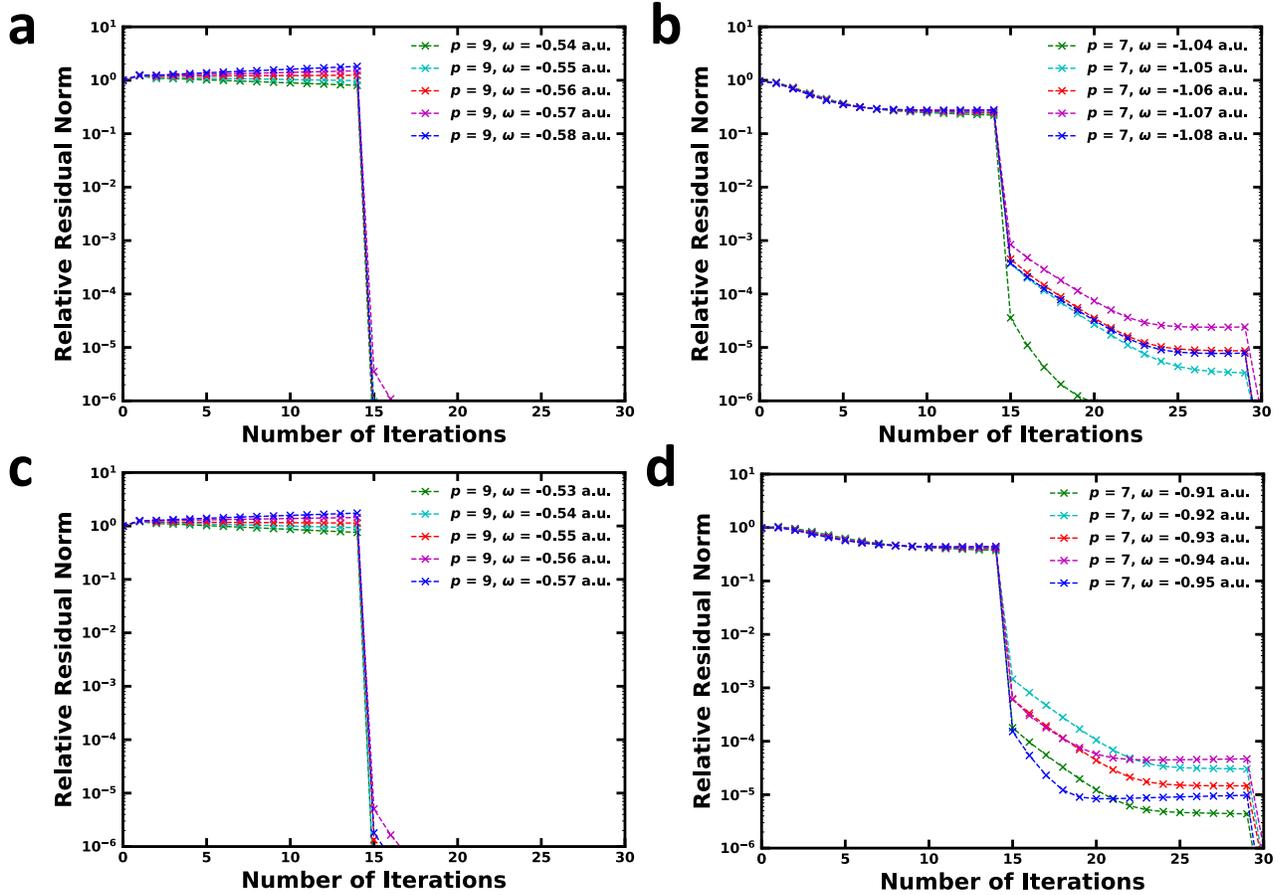}
\caption{Convergence performance test of the linear equation solver (DIIS adapted to complex space) used in the  calculations of ${\bf x}_p(\omega)$'s for an N$_2$ molecule at chosen orbital indices $p$'s and frequency domains. The number of iterates used in the test is 15. $\Delta\omega=0.01$ a.u. (0.272 eV) applies to all the chosen frequency domains. ({\bf a,c}) correspond to the GFCCSD and GFCC-i(2,3) calculations of the ${\bf x}_p(\omega)$'s attributing to the first main peak of the spectral function. ({\bf b,d}) correspond to the GFCCSD and GFCC-i(2,3) calculations of the ${\bf x}_p(\omega)$'s attributing to the first satellite peak of the spectral function.}
\label{conv}
\end{figure*}

Fig. (\ref{conv}) exhibits the convergence performance of our linear equation solver (DIIS adapted to complex space) in the frameworks of both GFCCSD and GFCC-i(2,3) methods for the calculation of ${\bf x}_p$'s over several frequencies that are close to the first main peak and first satellite peak in the spectral function of an N$_2$ molecule. 
The position of the first main peak is predicted to be at $\omega$=-0.558 a.u. (-15.185 eV) by GFCCSD and at $\omega$=-0.550 a.u. (-14.968 eV) by GFCC-i(2,3), respectively. The position of the first satellite peak is predicted to be at $\omega$=-1.058 a.u. (-28.792 eV) by GFCCSD and at $\omega$=-0.930 a.u. (-25.309 eV) by GFCC-i(2,3), respectively. Given the number of micro iterates to be 15, Fig. \ref{conv}a,c show the relative residual norm of ${\bf x}_p$ of the frequencies that are close to the main peak can be quickly reduced to be $<10^{-5}$ after one DIIS update in both the GFCCSD and GFCC-i(2,3) calculations, while the convergence of ${\bf x}_p$ of the frequencies domains that include the satellite peak is a bit slower (see Fig. \ref{conv}b,d), {\it i.e.} to reach the same accuracy as for the main peak, at least one more DIIS update would be needed in either GFCCSD or GFCC-i(2,3) calculation. 

This performance difference can be attributed to different peak features that impact the complication of solving the coupled linear equations (Eq. (\ref{newlineq})). Note that, different from the main peaks which are dominated by one-electron process, the satellite peaks have significant contributions from double excitations (two-electron process). Thus, in order to solve for ${\bf x}_p$, in the frequency domain for main peaks, the iterating update would be basically on the one-body part of ${\bf x}_p$ whose dimension is only $n_o$, while in the frequency domain for satellite peaks, the iterating update would be instead on the larger two-body part of ${\bf x}_p$ whose dimension is $n_o^2n_v$. Apparently, more variables would usually require more iterations to help the linear equations converge.


\section{results and discussion}

\begin{figure*}
\includegraphics[width=\textwidth]{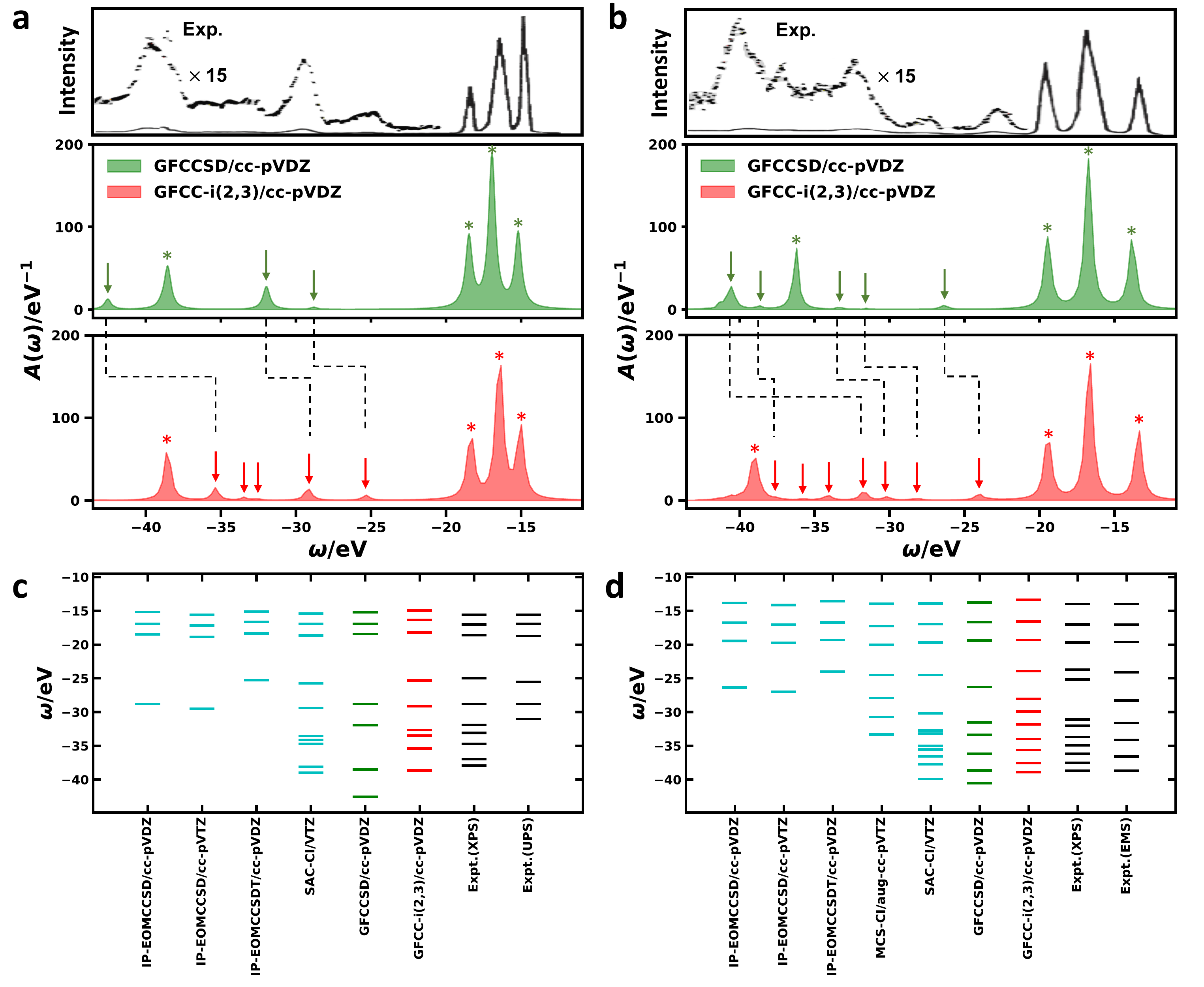}
\caption{Spectral functions, $A(\omega)$'s, of N$_2$ ({\bf a}) and CO ({\bf b}) molecules in the energy regime between -1.6 a.u. (-43.542 eV) and -0.4 a.u. (-10.886 eV) computed by the GFCCSD and GFCC-i(2,3) methods with $\eta=$ 0.01 a.u. (0.272 eV) and $\Delta\omega\le$ 0.01 a.u. ($\le$0.272 eV), with comparison with the peak positions extracted from the experimental spectra and other theoretical results ({\bf c,d}). In ({\bf a}) and ({\bf b}), main peaks (dominated by one-electron process) are labeled by asterisks, and satellite peaks are labeled by arrows. In ({\bf c}) and ({\bf d}), the IP-EOMCC results for the two molecules are from Ref. \citenum{hirata06_074111}, the symmetry-adapted-cluster configuration interaction (SAC-CI) results are collected from Ref. \citenum{ehara05_881}, the multichannel Schwinger configuration interaction (MCS-CI) results for the CO molecule are from Ref. \citenum{lebech12_094303}, and the experimental spectra and values are collected from Refs. \citenum{krummacher80_3993}, \citenum{krummacher83_1733} and \citenum{svensson91_184},
in which, the incident photoelectron energies ($h\nu$) were reported to be 50.3 eV and 55.1 eV, respectively. Note that the relative intensities between the different final ionic state manifolds exhibited in the photoelectron spectrum supposed to be different from those observed from the calculated spectral functions (see discussion in the text).
}
\label{co-n2-valence}
\end{figure*}

Our preliminary accuracy test of the GFCC-i(2,3) method is focused on the computed spectral functions of the N$_2$ and CO molecules, and their comparisons with the GFCCSD spectral function as well as other theoretical results and experimental observations. The experimental geometries\cite{herzberg2013} of the N$_2$ and CO molecules were used in all the calculations. The conventional ($\Lambda$-)CCSD calculations of the two molecules were performed first by using the NWChem suite of quantum chemical codes.\cite{nwchem} The converged $T$ and $\Lambda$ amplitudes, as well as the two-electron integral tensors, were then used by our pilot code to compute the IP-EOMCCSD type vector ${\bf x}_p(\omega)$ and the $\omega$-dependent retarded GFCC matrix elements. The spectral function is then given by the imaginary part of the retarded GFCC matrix,
\begin{equation}
A(\omega) = - \frac {1} {\pi} \text{Tr} \left[ \Im\left({\bf G}^{\text{R}}(\omega) \right) \right] 
= - \frac {1} {\pi} \sum_{p} \Im\left(G_{pp}^{\text{R}}(\omega) \right)~.
\end{equation}
The computed spectral functions the N$_2$ and CO molecules are shown in Fig. \ref{co-n2-valence}, and detailed information of the peaks are given in Tabs. \ref{tab1} and \ref{tab2}. The Hartree-Fock electronic configurations of the ground states of the N$_2$ and CO molecules can be described as $(1\sigma_g)^2$ $(1\sigma_u)^2$ $(2\sigma_g)^2$ $(2\sigma_u)^2$ $(3\sigma_g)^2$ $(1\pi_u)^4$ and $(1\sigma)^2$ $(2\sigma)^2$ $(3\sigma)^2$ $(4\sigma)^2$ $(1\pi)^4$ $(5\sigma)^2$, respectively. 

For both N$_2$ and CO molecules, 
it has been well-known that strong correlation effects exist in their inner-valence ionization process, and lead to the the breakdown of the MO picture. A typical example is the splitting of the single 2$\sigma_g$ line of N$_2$ (similarly 3$\sigma$ line of CO) into several satellite lines (see Ref. \citenum{schirmer77_149} for an early study). In the present study,
as shown in Fig. \ref{co-n2-valence}a,b, the GFCCSD/cc-pVDZ and GFCC-i(2,3)/cc-pVDZ calculations almost give the same main peak positions (labeled by asterisks) 
%
above -0.735 a.u. ($\sim$-20 eV). Below this energy regime, several satellite peaks started to emerge. Take N$_2$ for an example, the GFCCSD/cc-pVDZ calculation exhibits three satellite peaks between -43 eV and -20 eV which are dominated by two-hole-one-particle process, but still show significant 2$\sigma^{-1}$ configuration contributions. However, in comparison to previous experimental observations\cite{krummacher80_3993, krummacher83_1733, svensson91_184}, the positions of these satellite peaks are not well reproduced by the GFCCSD calculations. For the first satellite peak in the calculated spectral function of N$_2$ (similar in CO molecule case), the GFCCSD calculation predicted its position at $\sim$-28.79 eV, which is about 3 eV below the experimental value\cite{krummacher80_3993, svensson91_184}. For all the following satellite peaks, a systematic overestimation (blue shift) of the satellite peak positions can be observed from the GFCCSD results. On the other hand,
the GFCC-i(2,3) calculation significantly red-shifts the satellite peaks to be more close to the experimental observations. 
The largest red-shift for the N$_2$ molecule ($0.264$ a.u. or 7.184 eV) occurs to the $K~^2\Sigma_g^+$ state that is dominated by a $1\pi_u^{-1}2\sigma_u^{-1}1\pi_g$ two-electron process (see Tab. \ref{tab1}). For the CO molecule, the largest red-shift is $0.320$ a.u. (8.708 eV), and is associated with the $I~^2\Sigma^+$ state that is dominated by a $5\sigma^{-1}1\pi^{-1}1\sigma^\ast$ two-electron process (see Tab. \ref{tab2}). 
Besides, in comparison with the GFCCSD results, new satellite peaks are resolved from the GFCC-i(2,3) calculation. As shown in the spectral functions computed by the GFCC-i(2,3)/cc-pVDZ approach in Fig. \ref{co-n2-valence}a,b, despite of exhibiting relatively weak intensity, two new satellite peaks at -1.200 a.u. and -1.230 a.u. (that is, at -32.657 eV and -33.473 eV, assigned to the $G~^2\Pi_u$ and $I~^2\Sigma_u^+$ states, respectively, see Tabs. \ref{tab1}) can be easily observed for the N$_2$ molecule, and two new satellite peaks at -1.250 a.u. and -1.310 a.u. (that is, at -34.017 eV and -35.650 eV, assigned to the $J~^2\Sigma^+$ and $K~^2\Sigma^+$ states, respectively, see Tabs. \ref{tab2}) are observed for the CO molecule.

The significant red-shifts and emergence of these inner satellite peaks then distinguish the GFCCSD result and GFCC-i(2,3) result in terms of the agreement with the experimental results. As can be seen from the comparison of the peak positions obtained from the spectral function computed by the GFCC methods with those peak positions read from the experimental photoelectron spectra (Fig. \ref{co-n2-valence}c,d), the relative peak positions obtained from the GFCC-i(2,3)/cc-pVDZ calculation show better agreement with the experimental results\cite{svensson91_184,potts74_3}. For example, for the N$_2$ $C~^2\Sigma_u^+$ and CO $C~^2\Sigma^+$ ionic states, the GFCC-i(2,3)/cc-pVDZ results are $\sim$0.13 a.u. ($\sim$3.538 eV) and $\sim$0.09 a.u. ($\sim$2.449 eV) more close to the experiment results, respectively, than the GFCCSD/cc-pVDZ results.

It is worth emphasizing that, in Fig. \ref{co-n2-valence}, the comparisons of the computed spectral functions and the experimental photoelectron spectra are made in terms of peak positions, while the relative intensities between the different final ionic state manifolds exhibited in the photoelectron spectra are supposed to be different from those observed from the calculated spectral functions. 
This is because the intensity of the $n$-th ionic state in the photoelectron spectrum is given by\cite{cederbaum77_205, schirmer78_1889}
\begin{eqnarray}
P_{n} = \sum_{ij} \theta_{i,n} \theta^\ast_{j,n} \sum_{k} s_{k,i} s^\ast_{k,j} \delta(\epsilon_k + I_n - h\nu),
\end{eqnarray}
where $\epsilon_k$ is the kinetic energy of the emitted electron, $I_n$ is the $n$-th ionization potential, $h\nu$ is the energy of a photon beam, and $\delta(...)$ is Dirac delta function. 
$\theta_{i,n} = \langle \Phi_n^{N-1} | a_i | \Phi_0^{N} \rangle$
determines the relative intensity of of $n$-th ionic state in the spectral function for removing an electron from orbital $\phi_i$ ($|\Phi_0^{N}\rangle$ is the initial state of $N$-electron system and $|\Phi_n^{N-1}\rangle$ is the $n$-th state of ($N-1$)-electron system), and 
$s_{k,i} = \langle \phi_k | \mathbf{\hat{r}} | \phi_i \rangle$ is the dipole matrix element for photoionization ($\mathbf{\hat{r}}$ is the position operator, $\phi_k$ and $\phi_i$ are continuum one-particle state and bound one-particle state, respectively). In single particle picture (that is there is only one $\theta_{i,n}$ contributing significantly), the above equation can be simplified to 
\begin{eqnarray}
P_{n} = | \theta_{i,n} |^2 \sum_{k} | s_{k,i} |^2 \delta(\epsilon_k + I_n - h\nu),
\end{eqnarray} 
where $| \theta_{i,n} |^2$ is the so-called pole strength whose convoluted sums compose the spectral function. Therefore, $P_{n}$ is approximately proportional to $| \theta_{i,n} |^2$ only when $h\nu$ is much larger than $I_n$ (such that the dependence of $| s_{k,i} |^2$ on $\epsilon_k$ is weak) and there are no appreciable initial state correlation effects (i.e., in single particle picture). Also, as can be seen from Fig. \ref{co-n2-valence}, the computed spectral function include the effects of orbital degeneracy. Here, due to the degeneracy of $1\pi$ orbitals, the corresponding intensity at -16.300 eV for $N_2$ (-16.600 eV for CO) is almost twice as large as that corresponding to the ionization of the adjacent $\sigma$ orbitals.

In Fig. \ref{co-n2-valence}c,d, we also compare the positions of the peaks in the spectral function computed by our GFCC methods with other theoretical ionization potentials. As can be seen, early IP-EOM-CC calculations\cite{hirata06_074111} only reported the computed ionization potentials up to $\sim$-1.0 a.u. ($\sim$-27.214 eV)
Using the same cc-pVDZ basis set, the peak positions in this regime computed by the GFCCSD are almost same as the ionization potentials computed by IP-EOMCCSD method. By including the internal triple ($X_{p,3}(\omega)$) in the calculation, the peak positions from the GFCC-i(2,3) calculations are very close to those ionization potentials computed by IP-EOMCCSDT method. Beyond this regime, the GFCC-i(2,3) results, in comparison with the GFCCSD results, show better agreement with the SAC-CI results in terms of peak positions and associated configurations. The exploitation of larger basis set in the GFCC calculations would be expected to follow the trends observed in the IP-EOM-CC studies\cite{hirata06_074111} that larger basis sets  in the GFCC method would make the calculated results 
in the inner valence regime
gradually deviate from the experimental values. 

To have a more detailed examination, we found that, in the previous theoretical IP-EOM-CC and SAC-CI studies, a satellite peak classified as $D~^2\Pi$ state was also mentioned for the N$_2$ and CO molecules to be located slightly above the $C~^2\Sigma^+$ state. However, the intensity of the $D~^2\Pi$ state was reported to be extremely low. According to the previous SAC-CI study,\cite{ehara05_881} the ratio of the intensity of the $D~^2\Pi$ state to its adjacent satellite $C~^2\Sigma^+$ state is $\sim1/520$ for the N$_2$ molecule and $\sim1/13$ for the CO molecule. Given the already low intensity of the $C~^2\Sigma^+$ state, the direct observation of the $D~^2\Pi$ state from the computed spectral function would be hard (in Fig. \ref{co-n2-valence}a,b, the $D~^2\Pi$ state can not be clearly observed from the computed spectral function with $\Delta\omega=0.01$ a.u., or 0.272 eV). In fact, this state was originally even missing in the XPS spectra, and can only be identified in the UPS (He II) and DES spectra,\cite{svensson91_184} which favor different final states in comparison with the photoelectron spectra.

\begin{figure}
\includegraphics[width=0.5\textwidth]{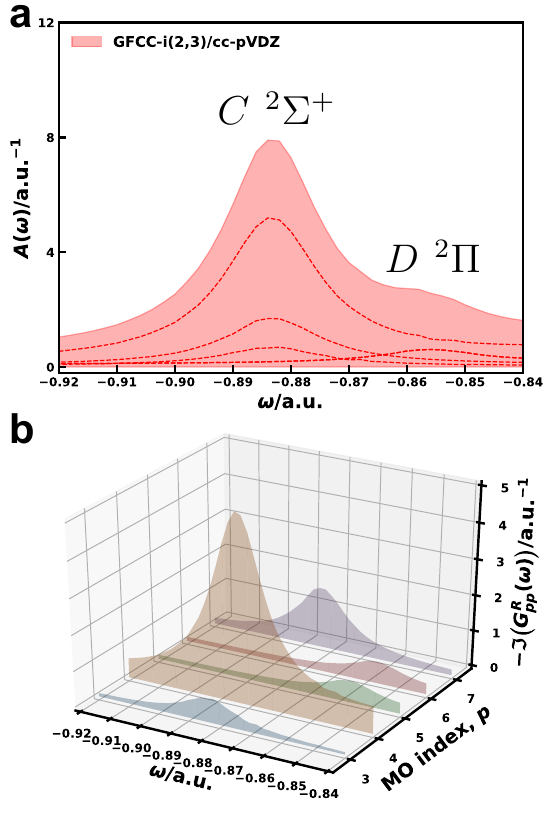}
\caption{({\bf a}) Spectral functions, $A(\omega)$'s, and ({\bf b}) its major contributions ({\it i.e.} the imaginary part of the diagonal Green's function matrix elements, $-\Im(G^R_{pp}(\omega))$) of CO molecule in the energy regime between -0.92 a.u. and -0.84 a.u. (that is, between -25.037 eV and -22.860 eV) computed by the GFCC-i(2,3) method with $\eta=$ 0.01 a.u. (0.272 eV) and $\Delta\omega=$ 0.002 a.u. (0.054 eV). }
\label{co}
\end{figure}
In order to resolve this state, we switch to a smaller frequency interval, $\Delta\omega=0.002$ a.u. (0.054 eV), in our GFCC calculations followed by a decomposition analysis. Fig. \ref{co} shows the computed spectral function of the CO molecule using the GFCC-i(2,3)/cc-pVDZ method in the energy regime of $\left[-0.92, -0.84\right]$ a.u. ($\left[-25.037, -22.860\right]$ eV) as well as the imaginary part of the individual diagonal Green's function matrix elements that significantly contributes to the computed spectral function in this regime. From Fig. \ref{co}a, slightly above the peak (at -0.88 a.u., or -23.948 eV) classified as the $C~^2\Sigma^+$ state, a small shoulder bump can be seen being centered at $\sim-0.856$ a.u. (-23.295 eV) From the decomposition analysis in Fig. \ref{co}b, we can further identify the bump as a satellite peak rather than a numerical fluctuation, and the major contributions to this shoulder satellite peak are from $-\Im(G^R_{55})$ and $-\Im(G^R_{66})$, which are associated with N$2p$ orbitals. This then confirm the classification of this satellite peak as the $D~^2\Pi$ state. The energy difference between the $C~^2\Sigma^+$ state and the $D~^2\Pi$ state in the GFCC-i(2,3), {\it i.e.} 0.024 a.u. ($\sim$0.653 eV), is slightly less than the experimental observation ($\sim$1.0 eV) and SAC-CI results ($0.88$ eV), and more close to the IP-EOM-CCSDT/cc-pVDZ result ($0.77$ eV). The main configurations of this satellite peak include $5\sigma^{-2}1\pi^\ast$ ($\sim$64\%), $4\sigma^{-1}5\sigma^{-1}1\pi^\ast$ ($\sim$21\%), and $5\sigma^{-2}2\pi^\ast$ ($\sim$6\%), which are consistent with MCS-CI and SAC-CI results.\cite{ehara05_881,lebech12_094303} 

Compared to the early Green's function/CI studies of the inner-valence ionization of N$_2$ and CO molecules\cite{schirmer77_149, ehara05_881, lebech12_094303, bagus1,bagus2}, the GFCC-i(2,3) calculation exhibits better agreement than the GFCCSD calculation in terms of overall spectral function profile (that is, the consistent prediction of the significant satellite peaks). Note that the previous Green's function study\cite{schirmer77_149} is able to provide more detailed information of the inner valence ionization by providing the positions of weaker satellite peaks which might not be observable in our present GFCC calculations. For example, in the same energy regime as used in Fig. \ref{co-n2-valence}, there were more than 10 ionization potentials (with corresponding Green's function residual $>$0.01) reported in Ref. \citenum{schirmer77_149}, while only six significant peaks observed from the GFCC-i(2,3) spectral function in the present study. As discussed in the Methodology section, this is because the previous Green's function methods (such as $2h$-$p$ TDA and ADC methods) are based on the diagonalization scheme which is in principle able to provide all the 1$h$ and $2h$-$p$ poles for the given theoretical level within the energy window of interest. The diagonalization scheme will be especially useful to resolve poles that have different characters but are lying very close to each other. In the present study, solving the linear equations and scanning over the energy points in the energy window of interest are on the other hand targeting the overall profile, and to explore more detailed information for a specific energy regime, a smaller energy interval and decomposition analysis of the peaks need to be employed, as evidenced from our above effort to identify the weak $D~^2\Pi$ state in the CO spectral function. However, even with the aid of a smaller energy interval and the decomposition analysis, it is still impossible to find dark states (corresponding to zero pole strength), and may still be difficult to find (or identify) the states with extremely low intensity. The missing of the states with low intensity would have direct impact on the accurate calculation of some properties (like vibronic coupling constant) in which the corresponding energies are indispensable. In such cases, it would be more reliable to combine the GFCC results with the results from recently proposed iterative subspace methods\cite{coriani12_1616, peng15_4146, zuev14_273} for the narrow energy window of interest to resolve all the possible states.

Since our GFCC approach is able to target any energy regime, we also give the core ionization potentials in Tabs. \ref{tab1} and \ref{tab2}. Compared to our previous GFCCSD/cc-pVDZ results\cite{kowalski18_4335}, the inclusion of internal triple in the GFCC framework does not the significantly change the main ionic peak positions, and both of the GFCCSD and GFCC-i(2,3) results show $\sim$3 eV deviation from the experimental observation\cite{bakke80_333, stohr13nexafs} and early fourth-order ADC study\cite{schirmer77_149, schirmer87_6031}. However, similar to what we observed in Ref. \citenum{kowalski18_4335}, employing larger basis set is able to reduce this deviation. For example, switching the cc-pVDZ basis set to Aug-cc-pVTZ basis, the position of the N $1\sigma^{-1}$ peak will be red-shifted to -15.09 a.u. ($\sim$-410.62 eV), which is much closer to the experimental value (the deviation is $<1$ eV). Note that the satellite peaks in the core regime is also an interesting topic, and there were efforts focusing on computing the satellite peaks accompanying the K-shell ionization of N$_2$ and CO molecules employing CVS-ADC(4) method\cite{schirmer77_149, schirmer87_6031}. However, the computation of the satellite peaks in the core regime is not covered in the present study, but will be one of the topics in our future work.

Finally, we would like to compare our GFCC-i(2,3) approach with other approximate EOM-CC (CCLR)-based methods where the treatment of triples are involved in different manner. Tab. \ref{tab3} lists the first three vertical ionization potentials of the N$_2$ and CO molecules computed by GFCC-i(2,3), CCSDT-n (n=1,2,3), and CC3 methods. Generally speaking, the CCSDT-3 and CC3 are able to provide the more accurate numbers than the CCSDT-n (n=1,2) and GFCC-i(2,3) methods, due to their more elaborate treatment of the triple equation, and the deviations from the experimental values for both methods are $<$0.1 eV. The quality of the GFCC-i(2,3) numbers lies between the CCSDT-3/CC3 and CCSDT-n (n=1,2), and the largest deviation in these GFCC-i(2,3) numbers is $\sim$0.3 eV (for N$_2$ $1\pi_u^{-1}$ state). Note that, both CCSDT-n (n=1,2) and GFCC-i(2,3) exclude the orbital relaxation associated with the $T_1$ cluster amplitude. In comparison to the CCSDT-n (n=1,2) which totally ignores the triple-single terms (that is the triple-single block of the associate Jacobian matrix is set to zero), the $(V_NT_2X_{p,1})_C$ term in the GFCC-i(2,3) approach contributes at lease to the second order correction to the triple equation. Since the GFCC-i(2,3) approach does not require $T_3$ cluster amplitude, its computational cost is even cheaper than the CCSDT-1 method.


\section{Conclusion}

In this paper, we propose a new category of the GFCC approximations, {\it i.e.} GFCC-i($n$,$m$) method, to improve the pole locations of the CC Green's function. In the GFCC-i($n$,$m$) method, the level of excitation ($m$) used to described the ionization potentials and electron affinities effects in the $N$$-$1 and $N$+1 particle spaces is higher than the level of excitation ($n$) used to correlate the ground-state coupled cluster wave function for the $N$-electron system. We demonstrate that in order to maintain size-extensivity of the GFCC matrix elements $m$ can not be larger than $n$+1. As our first practice in this direction, we implement the GFCC-i(2,3) method, in which the cluster operators $T$ and the de-excitation operator $\Lambda$ are kept at the CCSD level, while the auxiliary operators $X_p(\omega)$ and $Y_q(\omega)$ include the inner three-body terms (or internal triples) that can be obtained from one-body and two-body $X_p(\omega)$ and $Y_q(\omega)$ terms in an on-the-fly manner. The computational cost and storage requirement of the GFCC-i(2,3) method are significantly lower than the corresponding GFCCSDT method.
Our preliminary accuracy test has been focused on the accurate computing the spectral functions of the N$_2$ and CO molecules in the inner and outer valence regime. In comparison with the GFCCSD results, the computed spectral functions by the GFCC-i(2,3) method is greatly improved, and exhibit better agreement with the experimental results and other theoretical results. The improvement is particularly significant in terms of providing higher resolution of satellite peaks and more accurate relative positions of these satellite peaks with respect to the main peak positions. Similar strategy could also be applied to other CC methods. For example, it is feasible to combine the CCSD treatment of the ground state with excitation calculation that employs an extended biorthogonal CC expansion manifold comprising single ($1h$), double ($2h$-$1p$), and zero-th order triple ($3h$-$2p$) excitations. Such a treatment will be able to treat the one-hole ($1h$) main states through third order, and give a distinctly better description of the $2h$-$1p$ satellite states. Finally, it is worth mentioning that vibronic coupling is another important factor for accurately simulating spectral functions\cite{cederbau82_14, cederbaum07book}. Recent efforts include the development of similarity-transformed EOM vibrational coupled-cluster theory\cite{hirata18_054104}, and relevant studies for benzene and adenine molecules\cite{cederbaum88_2023, krylov10_12305}. Our future development will also be focusing on this direction.

\begin{table*}[h!]
\begin{center}
    \caption{Significant peaks in the spectral functions, $A(\omega)$'s, of N$_2$ molecule calculated in valence (including outer and inner valence) regimes by the GFCC-i(2,3) method with cc-pVDZ basis. For each peak, its relative intensity, significant Green's function matrix elements contributors, main configurations, as well as the weight of all the $|2h,p\rangle$ configurations (as a net effect of all contributing Slater determinants) are given.}
    \label{tab1}
\resizebox{2\columnwidth}{!}{
\begin{tabular}{l l l l l l l l l c c}
\\
\hline
& \multirow{2}{*}{\begin{tabular}[c]{@{}l@{}}State\\ (Symmetry) \end{tabular}} &  
	& Energy/eV&    
	& \multirow{2}{*}{\begin{tabular}[c]{@{}l@{}}Significant $G_{pp}^{\text{R}}$\\ (Contribution$\ge 0.05$) \end{tabular}} &  
	& \multirow{2}{*}{Main configurations ($|C|^2 \ge 0.02$)} 	
	& \multicolumn{2}{l}{Weight of} \\
& 	&  &                                                                                          
	&  &                                                                                                                   
	&  &  
	&\multicolumn{2}{l}{$|2h, p \rangle$'s} \\ \hline
& $X(^2\Sigma_g^+)$         		&  
	& -14.968        	&  
	& $G_{55}^{\text{R}}$(0.93)	&    
	& $3\sigma_g^{-1}$(0.91)        	&
	& 0.02        	\\[1ex]
& $A(^2\Pi_u)$      &  
	& -16.328           	&  
	& $G_{66}^{\text{R}}(0.48), G_{77}^{\text{R}}$(0.48)  	&  
	& $1\pi_u^{-1}$(0.94)           	&
	& 0.01        	\\[1ex]
& $B(^2\Sigma_u^+)$         		&  
	& -18.233           	&  
	& $G_{44}^{\text{R}}$(0.94)	&  
	& $2\sigma_u^{-1}$(0.90),	$1\pi_u^{-1}3\sigma_g^{-1}1\pi_g$(0.03)	&
	& 0.04	\\[1ex]
& $C(^2\Sigma_u^+)$   	&  
	& -25.309          	&  
	& $G_{44}^{\text{R}}$(0.95)		&  
	& $2\sigma_u^{-1}$(0.20), $1\pi_u^{-1}3\sigma_g^{-1}1\pi_g$(0.70),             &
	& 0.75	\\
&   	&  
	&      	&  
	& 	&  
	& $1\pi_u^{-1}3\sigma_g^{-1}2\pi_g$(0.06)            &
	& 	\\[1ex]
& $F(^2\Sigma_g^+)$     	&  
	& -29.119           	&  
	& $G_{33}^{\text{R}}(0.93)$, $G_{55}^{\text{R}}(0.05)$         	&  
	& $2\sigma_g^{-1}$(0.32), $3\sigma_g^{-1}$(0.04), $1\pi_u^{-1}2\sigma_u^{-1}1\pi_g$(0.46)                    &
	& 0.63	\\
&     	&  
	&      	&  
	&      	&  
	& $2\sigma_g^{-1}3\sigma_u^{-1}1\pi_g$(0.08), $1\pi_u^{-1}2\sigma_u^{-1}2\pi_g$(0.02)                    &
	& 	\\[1ex]
& $G(^2\Sigma_g^+,~^2\Sigma_u^+)$         	&  
	& -32.657           	&  
	& $G_{33}^{\text{R}}(0.16)$, $G_{44}^{\text{R}}(0.22)$,          	&  
	& $2\sigma_g^{-1}$(0.14), $2\sigma_u^{-1}$(0.03), $2\sigma_u^{-1}3\sigma_g^{-1}1\pi_g$(0.55),                     &
	& 0.79	\\
&          	&  
	&         	&  
	& $G_{66}^{\text{R}}(0.30)$, $G_{77}^{\text{R}}(0.30)$         	&  
	& $1\pi_u^{-1}3\sigma_g^{-1}1\pi_g$(0.17), $1\pi_u^{-1}2\sigma_u^{-1}1\pi_g$(0.02),              &
	&  	\\
&          	&  
	&         	&  
	&          	&  
	& $2\sigma_u^{-1}3\sigma_g^{-1}2\pi_g$(0.02)              &
	&  	\\[1ex]
& $I(^2\Sigma_g^+,~^2\Sigma_u^+)$     	&  
	& -33.473           	&  
	& $G_{33}^{\text{R}}(0.11), G_{44}^{\text{R}}$(0.84)		&  
	& $2\sigma_g^{-1}$(0.08), $2\sigma_u^{-1}$(0.08), $1\pi_u^{-1}3\sigma_g^{-1}1\pi_g$(0.68),                    &
	& 0.78	\\
&   	&  
	&      	&  
	& 	&  
	& $1\pi_u^{-1}2\sigma_u^{-1}1\pi_g$(0.02), $3\sigma_g^{-2}4\sigma_u$(0.03)            &
	& 	\\[1ex]
& $K(^2\Sigma_g^+)$             	&  
	& -35.378                      	&  
	& $G_{33}^{\text{R}}$(0.98)              	&  
	& $2\sigma_g^{-1}$(0.38), $1\pi_u^{-1}2\sigma_u^{-1}1\pi_g$(0.54),			&
	& 0.60	\\
&     	&  
	&	&  
	&      &  
	& $2\sigma_u^{-1}3\sigma_g^{-1}4\sigma_u$(0.02),			&
	& 	\\[1ex]
& $(^2\Sigma_g^+)$             	&  
	& -38.643                      	&  
	& $G_{33}^{\text{R}}$(0.99)              	&  
	& $2\sigma_g^{-1}$(0.85),	 $1\pi_u^{-1}2\sigma_u^{-1}1\pi_g$(0.08)		&
	& 0.13	\\[1ex]
& $(^2\Sigma_g^+,~^2\Sigma_u^+)$             	&  
	& -413.07                      	&  
	& $G_{11}^{\text{R}}$(0.46), $G_{22}^{\text{R}}$(0.54)              	&  
	& $1\sigma_g^{-1}$(0.44), $1\sigma_u^{-1}$(0.51)		&
	& 0.05	\\
\hline
\end{tabular}
}
  \end{center}
\end{table*}
\begin{table*}
\caption{Significant peaks in the spectral functions, $A(\omega)$'s, of CO molecule calculated in valence (including outer and inner valence) energy regimes by the GFCC-i(2,3) methods with cc-pVDZ basis. For each peak, its relative intensity, significant Green's function matrix elements contributors, main configurations, as well as the weight of all the $|2h,p\rangle$ configurations (as a net effect of all contributing Slater determinants) are given.}
\label{tab2}
\resizebox{2\columnwidth}{!}{
\begin{tabular}{l l l l l l l l l c c}
\\
\hline
& \multirow{2}{*}{\begin{tabular}[c]{@{}l@{}}State\\ (Symmetry) \end{tabular}} &  
	& Energy/eV&    
	& \multirow{2}{*}{\begin{tabular}[c]{@{}l@{}}Significant $G_{pp}^{\text{R}}$\\ (Contribution$\ge 0.05$) \end{tabular}} &  
	& \multirow{2}{*}{Main configurations ($|C|^2 \ge 0.02$)} 	
	& \multicolumn{2}{l}{Weight of} \\
&                        
	&  &                                                                                          
	&  &                                                                                                                   
	&  &  
	&\multicolumn{2}{l}{$|2h, p \rangle$'s} \\ \hline
& $X(^2\Sigma^+)$         		&  
	& -13.335        	&  
	& $G_{77}^{\text{R}}$(0.98)	&    
	& $5\sigma^{-1}$(0.96)        	&
	&  0.02	\\[1ex]
& $A(^2\Pi)$      &  
	& -16.600           	&  
	& $G_{55}^{\text{R}}(0.50), G_{66}^{\text{R}}$(0.50)  	&  
	& $1\pi^{-1}$(0.96)           	&
	& 0.02        	\\[1ex]
& $B(^2\Sigma^+)$         		&  
	& -19.322           	&  
	& $G_{44}^{\text{R}}$(0.97)	&  
	& $4\sigma^{-1}$(0.91)	&
	& 0.05	\\[1ex]
& $C(^2\Sigma^+)$   	&  
	& -23.948          	&  
	& $G_{33}^{\text{R}}$(0.09), $G_{44}^{\text{R}}$(0.65),		&  
	& $4\sigma^{-1}$(0.17), $5\sigma^{-1}$(0.05), $5\sigma^{-1}1\pi^{-1}1\pi^\ast$(0.62),             &
	& 0.70	\\
&    	&  
	&     	&  
	& $G_{77}^{\text{R}}$(0.21)		&  
	&  $5\sigma^{-1}1\pi^{-1}2\pi^\ast$(0.04)    	&
	& 	\\[1ex]
& $E,G(^2\Sigma^+,~^2\Pi)$     	&  
	& -28.030           	&  
	& $G_{33}^{\text{R}}(0.05)$, $G_{44}^{\text{R}}(0.06)$,         	&  
	& $3\sigma^{-1}$(0.06), $4\sigma^{-1}$(0.04), $1\pi^{-1}$(0.02),                     &
	& 0.67	\\
&   	&  
	& 	&  
	& $G_{55}^{\text{R}}(0.43)$, $G_{66}^{\text{R}}(0.43)$         	&  
	& $5\sigma^{-1}$(0.16), $4\sigma^{-1}5\sigma^{-1}1\pi^\ast$(0.12), 	&
	& 	\\
&   	&  
	& 	&  
	&     	&  
	& $1\pi^{-2}1\pi^\ast$(0.20), $5\sigma^{-2}1\pi^\ast$(0.04),	&
	& 	\\
&   	&  
	& 	&  
	&     	&  
	& $4\sigma^{-1}1\pi^{-1}1\pi^\ast$(0.05), $5\sigma^{-1}1\pi^{-1}1\pi^\ast$(0.17),	&
	& 	\\[1ex]
& $F(^2\Sigma^+)$         	&  
	& -29.935           	&  
	& $G_{33}^{\text{R}}(0.28)$, $G_{44}^{\text{R}}(0.34)$,          	&  
	& $3\sigma^{-1}$(0.02), $4\sigma^{-1}$(0.03), $5\sigma^{-1}$(0.08),                     &
	& 0.80	\\
&     	&  
	& 	&  
	& $G_{77}^{\text{R}}(0.33)$          	&  
	& $5\sigma^{-1}1\pi^{-1}1\pi^\ast$(0.59), $4\sigma^{-1}1\pi^{-1}1\pi^\ast$(0.10),    &
	& 	\\
&     	&  
	& 	&  
	&   	&  
	& $5\sigma^{-2}1\sigma^\ast$(0.02)    &
	& 	\\[1ex]
& $I(^2\Sigma^+)$    	&  
	& -31.840           	&  
	& $G_{33}^{\text{R}}(0.84), G_{44}^{\text{R}}$(0.14)		&  
	& $3\sigma^{-1}$(0.29), $4\sigma^{-1}$(0.03), $4\sigma^{-1}1\pi^{-1}1\pi^\ast$(0.46),                    &
	& 0.64	\\
&     	&  
	& 	&  
	&   	&  
	& $5\sigma^{-1}1\pi^{-1}1\pi^\ast$(0.04)    &
	& 	\\[1ex]
& $J(^2\Sigma^+)$             	&  
	& -34.017                      	&  
	& $G_{33}^{\text{R}}$(0.80), $G_{44}^{\text{R}}$(0.11),              	&  
	& $3\sigma^{-1}$(0.17), $4\sigma^{-1}1\pi^{-1}1\pi^\ast$(0.65),			&
	& 0.79	\\
&     	&  
	&   	&  
	& $G_{77}^{\text{R}}$(0.08)              	&  
	& $5\sigma^{-1}1\pi^{-1}1\pi^\ast$(0.03), $4\sigma^{-1}5\sigma^{-1}6\sigma$(0.02) &
	& 	\\[1ex]
& $K(^2\Sigma^+)$             	&  
	& -35.650                      	&  
	& $G_{33}^{\text{R}}$(0.33), $G_{55}^{\text{R}}$(0.30),              	&  
	& $3\sigma^{-1}$(0.27), $4\sigma^{-1}1\pi^{-1}1\pi^\ast$(0.03),		&
	& 0.64	\\
&  	&  
	& 	&  
	& $G_{66}^{\text{R}}$(0.30)              	&  
	& $5\sigma^{-1}1\pi^{-1}1\sigma^\ast$(0.54), $1\pi^{-2}1\pi^\ast$(0.02)	&
	& 	\\[1ex]
& $(^2\Sigma^+)$             	&  
	& -37.555                      	&  
	& $G_{33}^{\text{R}}$(0.83), $G_{33}^{\text{R}}$(0.09)              	&  
	& $3\sigma^{-1}$(0.31), $4\sigma^{-1}5\sigma^{-1}1\sigma^\ast$(0.50), $5\sigma^{-2}1\sigma^\ast$(0.03) 		&
	& 0.61	\\[1ex]
& $(^2\Sigma^+)$             	&  
	& -38.916                     	&  
	& $G_{33}^{\text{R}}$(0.99)              	&  
	& $3\sigma^{-1}$(0.85), $4\sigma^{-1}1\pi^{-1}1\pi^\ast$(0.04)		&
	& 0.14	\\[1ex]
& $(^2\Sigma^+)$             	&  
	& -298.26                     	&  
	& $G_{22}^{\text{R}}$(0.99)              	&  
	& $2\sigma^{-1}$(0.92)	&
	& 0.08	\\[1ex]
& $(^2\Sigma^+)$             	&  
	& -546.73                     	&  
	& $G_{11}^{\text{R}}$(0.99)              	&  
	& $1\sigma^{-1}$(0.97)	&
	& 0.03	\\
\hline
\end{tabular}
}
\end{table*}
\begin{table*}
\caption{Vertical ionization potentials of N$_2$ and CO molecules obtained from GFCC-i(2,3) and various EOM-CC (CCLR) based methods. The Aug-cc-pVTZ basis is used in all the calculations. The CCSDT-n and CC3 results are collected from Ref. \citenum{stanton99_8785}.}
\label{tab3}
\resizebox{1.2\columnwidth}{!}{
\begin{tabular}{l c l c l c l c l c l c l c l c}
\\
\hline
&& && &\multicolumn{7}{c}{CCSDT-n}& && & \\
\cline{6-12}
& State &  
	& GFCC-i(2,3) &    
	& n=1 &  
	& n=1b &
	& n=2 &
	& n=3 &
	& CC3 & 	
	& Exp. \\ \hline
& N$_2$ &  
	& &    
	& &  
	& &
	& &
	& &
	& & 	
	& \\	
& 2$\sigma_g^{-1}$ &  
	& 15.67 &    
	& 15.85 &  
	& 15.84 &
	& 15.89 &
	& 15.58 &
	& 15.54 & 	
	& 15.60 \\
& 1$\pi_u^{-1}$ &  
	& 17.28 &    
	& 17.35 &  
	& 17.34 &
	& 17.40 &
	& 16.95 &
	& 16.88 & 	
	& 16.98 \\
& 2$\sigma_u^{-1}$ &  
	& 18.91 &    
	& 19.02 &  
	& 19.02 &
	& 19.09 &
	& 18.87 &
	& 18.82 & 	
	& 18.78 \\	
& CO &  
	& &    
	& &  
	& &
	& &
	& &
	& & 	
	& \\	
& 5$\sigma^{-1}$ &  
	& 14.17 &    
	& 14.17 &  
	& 14.15 &
	& 14.25 &
	& 13.94 &
	& 13.86 & 	
	& 14.01 \\
& 1$\pi^{-1}$ &  
	& 17.09 &    
	& 17.33 &  
	& 17.35 &
	& 17.36 &
	& 17.10 &
	& 17.09 & 	
	& 16.91 \\
& 4$\sigma^{-1}$ &  
	& 19.81 &    
	& 19.88 &  
	& 19.88 &
	& 19.91 &
	& 19.77 &
	& 19.76 & 	
	& 19.72 \\	
\hline
\end{tabular}
}
\end{table*}
%


\section{acknowledgement}

This work was supported by the Center for Scalable, Predictive methods for Excitation and Correlated phenomena (SPEC), which is funded by the U.S. Department of Energy (DOE), Office of Science, Office of Basic Energy Sciences, the Division of Chemical Sciences, Geosciences, and Biosciences.
All calculations have been performed using the Molecular Science Computing Facility (MSCF) in the Environmental Molecular Sciences Laboratory (EMSL) at the Pacific Northwest National Laboratory (PNNL). EMSL is funded by the Office of Biological and Environmental Research in the U.S. Department of Energy. PNNL is operated for the U.S. Department of Energy by the Battelle Memorial Institute under Contract DE-AC06-76RLO-1830. B. P. acknowledges the Linus Pauling Postdoctoral Fellowship from PNNL. B.P. and K.K. acknowledge the constructive comments from the reviewers.

\bibliography{gfcc}

\end{document}